%% file: Casertano_multi.tex
\documentstyle[psfig,preprint]{aastex}

\input psfig

\def\package#1{{\bf #1}}
\def\task#1{{\bf #1}}   
\newcommand\variance{{\rm Var}}
\newcommand\etal{et al.}
\newcommand\sextractor{SExtractor}
\def\avg#1{\langle #1 \rangle}
\newcommand\digit{$\phantom{0}$}
 
\begin{document}

\input Casertano_main.tex

\clearpage

\input Casertano_realfigures.tex

\clearpage 

\input Casertano_tables.tex

\end{document}

%% file: Casertano_main.tex
\slugcomment {To appear in the December 2000 issue of {\it The
Astronomical Journal}}

\shorttitle {WFPC2 Observations of the HDF-S}
\shortauthors {S.~Casertano et al.}

\title {WFPC2 Observations of the Hubble Deep Field-South}

\author {Stefano~Casertano\altaffilmark{1,2}, 
Duilia~de Mello\altaffilmark{1},
Mark~Dickinson\altaffilmark{1},
Henry~C.~Ferguson\altaffilmark{1}, 
Andrew~S.~Fruchter\altaffilmark{1},
Rosa~A.~Gonzalez-Lopezlira\altaffilmark{1},
Inge~Heyer\altaffilmark{1},
Richard~N.~Hook\altaffilmark{3},
Zolt Levay\altaffilmark{1},
Ray~A.~Lucas\altaffilmark{1},
Jennifer~Mack\altaffilmark{1},
Russell~B.~Makidon\altaffilmark{1},
Max~Mutchler\altaffilmark{1},
T.~Ed~Smith\altaffilmark{1},
Massimo~Stiavelli\altaffilmark{1,2},
Michael~S.~Wiggs\altaffilmark{1},
Robert~E.~Williams\altaffilmark{1}
}

\altaffiltext{1}{Space Telescope Science Institute, 3700 San Martin Drive,
Baltimore, MD 21218, USA}
\altaffiltext{2}{On assignment from the Space Sciences Division of the
European Space Agency}
\altaffiltext{3}{ESO/ST-ECF, Karl-Schwartzschild-Strasse 2, D--85748, Garching, Germany}

\begin{abstract}

The Hubble Deep Field-South observations targeted a
high-galactic-latitude field near QSO J2233-606.  We present WFPC2
observations of the field in four wide bandpasses centered at roughly
300, 450, 606, and 814 nm.  Observations, data reduction procedures, and
noise properties of the final images are discussed in detail.  A catalog
of sources is presented, and the number counts and color distributions
of the galaxies are compared to a new catalog of the HDF-N that has been
constructed in an identical manner.  The two fields are qualitatively
similar, with the galaxy number counts for the two fields agreeing to
within 20\%.  The HDF-S has more candidate Lyman-break galaxies at $z >
2$ than the HDF-N.  The star-formation rate per unit volume computed
from the HDF-S, based on the UV luminosity of high-redshift candidates,
is a factor of 1.9 higher than from the HDF-N at $z \sim 2.7$, and a
factor of 1.3 higher at $z \sim 4$. 

\end{abstract}


\section {Introduction}

The Hubble Deep Field-South (HDF-S) consists of a large set of
observations of an otherwise unremarkable field around the QSO J2233-606
($ z = 2.24 $), taken in parallel by three instruments aboard the Hubble
Space Telescope (HST): the Wide Field and Planetary Camera 2 (WFPC2),
the Space Telescope Imaging Spectrograph (STIS), and the
Near Infrared Camera and Multi Object Spectrometer (NICMOS).  This QSO
was chosen because, among other characteristics, it is situated within
the southern HST Continuous Viewing Zone (CVZ), a narrow declination
range where, at specific times during the year, HST can make
observations uninterrupted by Earth occultations.

The majority of the observations were taken at a fixed pointing (the
``main field''), with STIS aimed directly at the QSO, and WFPC2 and
NICMOS imaging areas a few arcminutes away.  The observations are
described in a series of papers: Williams {\etal} (2000) details the
field selection and the overall strategy, Fruchter {\etal} (2000) the
NICMOS observations, Gardner {\etal} (2000) the STIS imaging
observations, and Ferguson {\etal} (2000) the STIS spectroscopic
observations.  The present paper is concerned with the WFPC2
observations of the main field.  Less deep observations were obtained of
selected contiguous regions around the main field (the ``flanking
fields''), and these are described in Lucas {\etal} (2000). 

The WFPC2 main field observations were centered around RA \hbox{$ \rm
22^h \, 32^m \, 56\fs 22$}, Dec \hbox{$-60^\circ 33' 02\farcs 69$}
(J2000), about $ 5' $ West of the QSO on which the STIS observations
were centered, and consist of a total of approximately 450 ks worth of
exposures in the four filters used for the original Hubble Deep Field in
the Northern hemisphere (HDF-N, Williams {\etal} 1996): F300W, F450W,
F606W, and F814W.  The rationale underlying this filter choice is
explained in detail in Williams {\etal} (1996).  Briefly, these filters
offer as broad a total wavelength coverage as can be obtained with
reasonable depth with WFPC2, while also providing color information. 
Their width (about 30\%) ensures maximum sensitivity within each
wavelength range, while the total wavelength interval thus covered
permits the identification of galaxies over an interesting range of
redshifts, corresponding to the Lyman limit crossing each of the bluer
filters.  The photometric constraints on redshifts allow an estimate of
the number of star-forming galaxies as a function of redshift to depths
greater than can be achieved with direct spectroscopic techniques, and
have been used to constrain the global star formation history of the
Universe (e.g., Madau {\etal} 1996).  At the same time, the availability
of deep images in different wavebands has allowed comparative studies of
the morphology and size in various filters, and of how they relate to
other properties of the galaxies (e.g., Abraham {\etal} 1996).  Perhaps
more importantly, the existence of a common deep field has helped focus
the redshift-gathering efforts with large telescopes to present a deeper
view of the distribution of galaxies than could have been achieved by
the effort of a single group. 

The WFPC2 observations of HDF-S have been designed to be as similar as
possible to those of the HDF-N, to permit direct comparisons
between the two fields and to provide a similar point of focus for large
telescopes now available in the Southern hemisphere.  As with the Hubble
Deep Field, they have been made public after a short period of time,
during which only basic data reduction was carried out.  Because of the
higher complexity of the observations, described further in Section~2,
and of the involvement of three separate instruments, the data
processing took place over a somewhat longer period of about six weeks. 
The observations took place between September 29 and October 10, 1998;
both raw and the Version 1 processed data were made public on November
23, 1998.  Some refinements were subsequently applied to the data
processing, and a Version 2 of the WFPC2 images has been made public in
June 1999.  This paper describes the Version 2 data products,
highlighting the improvements over Version 1 where appropriate. 

\section {The Observations}

\subsection {Scattered Light Background and Observation
Scheduling}\label{sec:scheduling_strategy}

As was the case for the HDF-N, the WFPC2 observations were
carefully scheduled to take into account the varying sky background
during each HST orbit.  While observing in the CVZ, HST points within
about $ 20^\circ $ of the surface of the Earth.  When on the day side,
the illuminated Earth significantly increases the sky background, with
this increase becoming greater as the telescope pointing approaches the
illuminated Earth limb.  The observations in F450W, F606W and F814W are
essentially sky-limited, and thus are much less efficient when the
illuminated side of the Earth is visible from HST.  On the other hand,
observations in F300W are closer to being read-noise limited, and thus
suffer less from an elevated sky background.  As the sky background is a
predictable function of orbital phase, the observations were scheduled
so that F300W was used during the periods of high background, and the
other filters during periods of normal background.  For more details,
see Williams {\etal} (2000). 

\subsection {Dithering Strategy}

The HDF-N utilized a dithering strategy in which observations were
obtained at nine separate pointings separated by about $ 2'' $.  The
shifts between pointings had a double advantage.  First, by overlaying
slightly different areas of the detector within each point on the sky,
they allowed the averaging and smoothing of low-level detector features,
such as slight imperfections in the flat fields and dark current
subtraction, hot pixels, etc, resulting in more uniform images.  Second,
since the pointings were at different fractional pixel positions, it was
possible to reconstruct a combined image with better sampling and
quality than the original, undersampled WFPC2 images, thus greatly
increasing the ability to discern small-scale details.  The ``drizzle''
software package (Fruchter \& Hook 1997, 1999) was developed in order to
perform the necessary image reconstruction efficiently and reliably.  On
the other hand, the HDF-N observations were also designed to have
multiple observations in each filter taken at precisely the same
pointing, so that the abundant cosmic rays could be rejected efficiently
using standard techniques. 

The HDF-S situation was markedly different.  The ability to dither the
observations was limited by the requirements of STIS spectroscopy and
the need to keep the QSO within the aperture.  Long slit observations
prohibited translations perpendicular to the slit; echelle observations,
which use one of five small apertures, limited the ability to translate
in either direction.  In consequence, our dithering strategy relied
heavily on changing the telescope roll angle, resulting in different
shifts for different parts of the camera.  Thus, a consistent subpixel
dithering strategy was not feasible.  In addition, optimal NICMOS
observations required a variety of shifts, including some large
ones (up to $ 7 '' $); with such large shifts, the geometric distortion
in WFPC2 also prevented a consistent subpixel dithering strategy. 

Therefore we adopted a diametrically opposite strategy.  Being unable to
create a pointing strategy resulting in uniform offsets for each
location within the field of view, we effectively ``randomized'' the
pointings, using a large number of different shifts and rolls---as many
as the STIS constraints allowed.  The consequence was a quasi-random set
of subpixel shifts in the pointing, allowing a near-uniform coverage of
each pixel throughout WFPC2.  Some examples of total and fractional
(subpixel) shifts are shown in Williams {\etal} (2000).

The obvious drawback of this strategy was that multiple images at each
pointing were not always available for direct cosmic ray rejection. 
Instead we had to rely on new capabilities within the {\package{dither}}
and {\package{ditherII}} STSDAS packages (Fruchter {\etal} 1997) which
allow cosmic ray rejection in cases where all the images have different
pointings.  The procedure worked very well, although it did increase the
processing complexity. 

\subsection {Observing Log}

The list of observations, with their time, length, planned pointing, and
quality flags is given in Table~\ref{table:obslist}.  The various quality flags are
explained further in Section 3. 

\section {Data Reduction}

All observations obtained for the WFPC2 main field were processed with a
procedure that included the following steps: preliminary quality
checking, pipeline processing, hot pixel removal, preliminary image
registration, preliminary combination, final image registration, cosmic
ray rejection, interchip alignment, sky background correction, scattered
light correction, final combination, and astrometric calibration. 

Lacking matched pairs (same filter, same pointing) for the majority of
WFPC2 exposures, cosmic ray rejection was carried out by taking
advantage of the new capabilities of the dithering software.  This
cosmic ray rejection is based on a three-stage process.  Each individual
input image is registered and separately drizzled to the same output
frame.  These individual images are not summed, but are combined using a
robust median rejection technique to produce an intermediate image with
somewhat lower resolution, but clean from cosmic rays.  The intermediate
image is then projected back to the frame of each input image in turn,
and compared to identify and mask cosmic rays, bad pixels, and other
imperfections.  Finally, the masked images are combined using the
standard {\package{dither}} at the full resolution and sampling allowed
by the data.  We also included extra steps to verify and improve the
measurement of the pointing for individual images. 

Because of the relatively large pointing shifts (up to $ 13'' $) and
chip rotations, we decided, unlike the HDF-N, to combine all
images into a single mosaic comprising all four WFPC2 chips.  This
reduces the data loss which would otherwise occur because parts of the
sky switch between chips when the relatively large pointing shifts
occur, and it provides the community with a single image for each filter
rather than the four separate images (one per chip) produced for the
HDF-N.  Processing all the data for each filter into a single
final image was slightly more complex, as it required careful chip
realignment to avoid mismatches in positions in the overlap regions and
to produce an astrometrically correct, nearly seamless final image. 
Because of the dithering strategy mandated by the STIS constraints, the
seams running roughly in the E-W direction---between WF3 and WF4 and
between PC and WF2---largely disappear in the final image, whereas the
seams running N-S---between WF2 and WF3 and between WF4 and PC---are not
as well covered, and there is a noticeable drop in signal-to-noise ratio
between these chips. 

\subsection {Calibration and Quality Verification}

\subsubsection {Quality Control}

As a first step, each image went through a direct inspection and quality
control.  Any obvious artifacts (satellite trails, scattered Earth
light, cosmic ray clusters, charge trails left by previous images) were
noted.  About 30\% of the images were flagged in some way.  Many of the
defects, such as satellite trails, could be corrected with normal
processing; others required masking the affected area by hand.  Bias
jumps and scattered Earth light were treated in special steps described
below.  Quality flags set during this process are noted in
Table~\ref{table:obslist}, with an indication of how the problem was
addressed, if appropriate, and for each case we have verified after the
fact that the anomaly was corrected satisfactorily. 

\subsubsection {Pipeline Processing}

The standard pipeline processing was used to obtain calibrated images. 
A new set of reference files (superbias, superdark, flat fields) was
obtained shortly before the HDF-S observations as part of the normal
WFPC2 update process.  The new reference files have since become
standard for WFPC2 processing. 

The new superbias file, named I9817383U.R2H, was generated in August
1998 from an average of 120 on-orbit bias frames taken between December
4, 1997 and August 13, 1998.  Each bias image was corrected for its mean
bias level, using separate averages for even and odd columns as is
standard for WFPC2 images.  The resulting images were combined using the
STSDAS task {\bf mkdark}, which rejects cosmic rays via an iterative
process.  Four passes were used, with rejection levels set at $ 6, 5, 5,
$ and $ 4 \sigma $ respectively.  The task also rejects pixels adjacent
to each cosmic ray that differ from their average.  Pixels with fewer
than 100 images contributing were flagged as ``bad''. 

The new superbias does not differ in any major way from the previous
WFPC2 superbias files, although it is generally advisable to use the
superbias closest to the date of the observations. 

The new superdark frame, named I9T1701QU.R3H, was produced in September
1998 from 120 dark images obtained between May 11 and August 21, 1998. 
Each dataset was processed by removing the mean bias level and then the
new superbias image.  Files were then combined using the STSDAS task
{\bf crreject} with rejection levels set at $ 4, 3, $ and $ 2 \sigma $. 
Pixels with fewer than 60 contributors were flagged as ``bad''. 
Similarly, pixels with a high dark current ($ > 0.02 $ DN/s) were
flagged as ``hot''.  The new superdark reflects the gradual increase in
WFPC2 dark current over the years; the total dark signal was about 40\%
higher at the time of the HDF-S observations than at the time of the
HDF-N observations.  In the vast majority of observations,
however, noise due to dark current is a minor contribution to the total
noise, and thus the increase in dark current does not significantly
affect the sensitivity of the observations. 

The new flat field reference files, named I9T1701IU.R4H (F300W),
I9T1701KU.R4H (F450W), I9T1701MU.R4H (F606W), and I9T1701OU.R4H (F814W),
were produced in September 1998 using new Earth flats obtained over the
preceding year.  They differ from previous flats primarily because of
features at the pyramid edge, which have shifted slightly in pixel
coordinates.  The shift, approximately 1 pixel, is a consequence of
slight motions of the individual detectors within the WFPC2 instrument;
these motions are also reflected in the relative positions of the
detectors on the plane of the sky (cf.~Section~\ref{sec:interchip}).  A
very small residual background variation, probably related to slight
changes in the interchip alignment, was removed empirically in the sky
background correction phase (cf.~Section~\ref{sec:skyflattening}). 

The pipeline processing consists of the following steps: masking of
known bad pixels; analog-to-digital conversion; subtraction of the bias
level, separately for even and odd columns; subtraction of the superbias
image; subtraction of the superdark image, scaled by the nominal dark
time of the observation; and multiplication by the (inverse) flat field
for the appropriate filter.  This process, documented in the HST Data
Handbook, Version 3 (Voit {\etal} 1997), results in high-quality
calibrated images with background and pixel response usually constant to
better than $ 1\% $.  The full pipeline processing was carried out in
the standard way using the new reference files. 

\subsubsection {Hot Pixel Removal}

Hot pixels, i.e., pixels with elevated dark current, are formed in WFPC2
on a daily basis, as a result of the continual impinging of charged
particles onto the detectors.  Approximately 30 new hot pixels appear
every day in each detector (WFPC2 Handbook).  The majority of these hot
pixels disappear each month when the detectors are warmed up to room
temperature to remove contaminants from the optical surfaces. 

Most hot pixels are masked and excluded as part of the cosmic ray
removal process, since the image dithering moves each hot pixel onto
different areas of the image.  However, some hot pixels are not ``hot''
enough to hit the threshold for cosmic ray removal, especially when they
fall near an object.  Therefore prior identification of hot pixels is
desirable whenever possible. 

In order to track the increasing population of hot pixels, we produced
daily hot pixel masks by median combination of all F300W images taken in
a given day.  The F300W images are generally well-suited for this
purpose, since they contain few objects and have low overall counts. 
The median combination excludes astronomical objects which fall on different places
on the chip due to the changes in pointing, and makes hot pixels clearly
discernible.  These are then flagged in the daily hot pixel mask. 
Images taken each day were combined with the appropriate daily mask, so
that each hot pixel was masked.  Insufficient images were available for
October 4, 9 and 10; we used the October 5 mask for October 4 and the
October 8 mask for October 9 and 10. 

\subsection {First Pass: Preliminary Combination}

The first combination of the data was aimed at producing an image that
could be used to verify the registration of individual exposures and to
identify cosmic rays and other imperfections.  For this purpose, the
emphasis of the first combination was on robustness, rather than on
obtaining the best possible signal-to-noise ratio or resolution.  The
first combined image had a pixel scale of 60 milli-arcseconds (mas),
more than adequate for a proper cosmic ray rejection, but not as
demanding in terms of number of input images and pointing quality as the
scale of 40 mas/pixel desired for the final image.  The first
combination included determination of preliminary registration,
median-combination of ``good'' images, and final image registration. 

\subsubsection {Preliminary Image Registration}

Although robust with respect to occasional deviations, the preliminary
combination needed a good-quality typical image registration, to about a
quarter of a pixel, to avoid interimage blurring of sources and improper
cosmic ray rejection.  This preliminary registration was performed using
two separate methods.  The first was to adopt the position recorded in
the jitter file, derived from the average measured position of the guide
stars throughout each exposure.  The second was to cross-correlate each
image with a cosmic-ray cleaned image pair chosen as reference.  (The
second method could not be used for F300W because of the low signal
level.) The images for which the two methods gave discrepant results
(about 20\%) were rejected from the first combination, and were later
registered with respect to the others by cross-correlation with the
median-combined image. 

\subsubsection {The Median-Combined Image}

After excluding all images with inconsistent registration, we used the
remaining images to produce a single, median-combined mosaic image for
each filter.  The combination entailed the following steps:

\begin {itemize}

\item For each input image, {\task{drizzle}} was used to produce a
mosaic of the four chips (this mosaic had a small inter-chip gap); the
registration information was used to line up all images into a common
reference position which was chosen to have North approximately up.  The
{\task{drizzle}} parameters were chosen to have a pixel size of 60~mas
and a footprint of 100~mas, thus producing a modest additional blurring
of the input images.  This blurring, which affected each input image in
the same way, helped in the identification of cosmic rays by smoothing
out apparent variations in source signal due to the undersampling of
WFPC2. 

\item The mosaic images were then median-combined, excluding all pixels
marked ``bad'' in the data quality file, ``hot'' in the daily hot pixel
list, or too close to the edges of the chip or of the pyramid beam splitter. 

\end {itemize}

As a result, we obtained a coregistered, largely cosmic-ray free image
with a sampling of 60~mas and a slightly larger PSF than each input
image.  Some cosmic rays remained in the regions where fewer than three
exposures could be combined, because these regions lack sufficient
information to distinguish between objects and cosmic rays. 

\subsection {Other Adjustments}

In preparation for the second and final pass, the preliminary combined
image was used to improve the image registration, to identify cosmic
rays on the original images, and to refine the interchip alignment.  The
images were also photometrically balanced to compensate for zero point
differences between the detectors. 

\subsubsection {Improved Image Registration}

Individual images taken in F450W, F606W, and F814W were re-registered
with respect to the median image by cross-correlating each individual
mosaic with the median image.  Cross-correlating the mosaics, rather
than each chip separately, yields a more accurate and more robust
measurement of the central position of the image, which is especially
helpful in the F450W images with their lower signal level. 

In a few cases, the quality of the cross-correlation indicated a
possible anomalous rotation.  For these cases we determined the image
rotation angle by drizzling the individual image with various rotations
and then cross-correlating different image sections with the preliminary
combined image.  The correct rotation angle was determined by requiring
that the measured shifts in both coordinates be the same for all image
sections. 

Images in F300W required a different procedure due to the low signal
level that caused even the mosaic cross-correlation to be dominated by
cosmic rays.  We cut out two 100-pixel regions around the two brighest
stars, and cross-correlated those against the reference image.  The
shifts were generally in good agreement and their average was adopted;
in a few cases, one of the stars was affected by a cosmic ray and the
other was used.  For one image, both stars were affected by a cosmic ray
and unusable; that image, identified in the Observation Log, was
excluded from the final combination. 

The formal uncertainties in the registration obtained for individual
exposures are typically 5--10 mas for F450W, F606W, and F814W, and
10--20 mas for F300W. 

\subsubsection {Cosmic Ray Rejection}\label{sec:crrej}

A special feature of the {\package{dither}} package was exploited to
identify and flag cosmic rays in all WFPC2 images, a process that can be
difficult in the absence of cosmic-ray splits.  The conceptual basis of
the cosmic ray rejection procedure is to compare each individual
exposure to a ``predicted'' exposure based on the median-combined image. 
The predicted exposure was obtained via the task {\task{blot}}, which
computes the expected flux in each pixel of each original exposure using
the median-combined image, taking into account the geometric distortion
of the WFPC2 cameras.  The process is conceptually the reverse of
{\task{drizzle}}, hence the name. 

The predicted image produced by {\task{blot}} is expected to be somewhat
different from the true exposure because of the undersampling of the
WFPC2 images and of the additional blurring due to the median
combination itself.  In order to determine whether the predicted and
true images are consistent, the {\package{dither}} package uses a
special algorithm.  For each pixel, the task {\task{deriv}} determines
the largest difference (in absolute value) between that pixel and each
of the neighboring pixels, and the task {\task{driz\_cr}} compares this
value with the difference between predicted and actual value.  This
process was in general very successful except close to the edges of the
median image, where there was not enough information to identify cosmic
rays properly.  Pixels affected by cosmic rays were identified in the
cosmic ray mask and excluded from the final combination. 

\subsubsection {Interchip Alignment}\label{sec:interchip}

The relative alignment of the WFPC2 detectors has changed slowly over
time.  The astrometric solutions derived by Holtzman {\etal} (1995) and
by Gilmozzi {\etal} (1996) were very accurate when they were determined,
but there is now evidence that the interchip alignment has changed by as
much as $ 0\farcs15 $ over the last three years, probably because of a
slow change in the WFPC2 optical bench.  Therefore it was necessary to
redetermine the interchip registration of WFPC2 for the HDF-S
observations.  Some of the pointing changes were large enough to cause
significant areas of the sky to be covered by different chips in
different exposures.  The image registration described above ensured
that the central regions of each detector would be properly aligned from
image to image, thus we could use the overlap regions to verify the
relative alignments. 

The process we used was the following.  a) For each filter, a separate
image was produced for each of the detectors.  These images overlap by
about $ 10'' $ in the N-S direction, and by about $ 3'' $ in the E-W
direction due to the variety of pointings used.  b) The positions of
objects in the overlap area were compared to determine shift corrections
with WF4 used as the reference; for the PC we used the average of the
shift determined by comparison with WF2 and WF4.  c) In addition, we
also determined the very small inter-filter shift due to both the
so-called filter wedge and the difference in the position of the
reference image. 

We then corrected the shifts used by {\task{drizzle}} for each detector
to compensate for the differences we found, keeping WF4 in F450W as the
reference.  After these updates, we estimate the relative alignment of
the images in different filters to be good to better than 5 mas, and
between detectors to about 15 mas. 

\subsubsection {Photometric balancing}

The four WFPC2 detectors are known to have slightly different
sensitivities.  Prior to any further processing, we rescaled each image
to compensate for the difference in sensitivity between detectors,
using the WF3 as reference.  Thus, the published zero points for WF3
apply to the final combined image.  See Table~\ref{table:photzpt} for
more details. 

\subsection {Other Corrections}

A number of additional small corrections were applied to the individual
images in order to improve the quality of the final combined image. 
While none of these corrections made a substantial impact on the
scientific results of the HDF-S project, they contributed to improve the
depth, flatness, and cosmetic appearance of the final image.  These
corrections include: a better sky background subtraction, a correction
for the scattered Earth light, and special processing to remove bias
jumps, satellite trails, and residual images. 

\subsubsection {Sky background correction}\label{sec:skyflattening}

A very small, but detectable systematic variation of the sky brightness
was apparent in the flat-fielded images in F606W and F814W.  This
variation was probably due to subtle changes in the flat field, possibly
related to the slow relative drift of the WFPC2 detectors, and produced
a slightly mottled combined image.  This residual sky background was
efficiently removed by the following procedure.  (1) Pipeline-calibrated
images were scaled to unit sky and median combined without shifting. 
Because of the extensive dithering, this effectively removes all
sources.  (2) A slowly-varying function was fitted to the sky image (a
Chebyshev function of order 3 was used).  (3) This fitted function was
scaled to the observed mode of the sky in each image and subtracted. 
The use of this procedure improves the sky flatness by about 0.4\%, a
very small gain that is nonetheless noticeable especially in F606W and
F814W. 

\subsubsection {Scattered light correction}

Images taken during daytime have an elevated background due to light
from the illuminated Earth, scattered off the support structure of the
HST secondary.  Besides the higher background, these images also suffer
from a distinctive cross-shaped pattern (the ``Earth cross'') due to the
shadowing of the HST secondary by the support structure of the WFPC2
repeater.  Our observing strategy
(Section~\ref{sec:scheduling_strategy}) ensured that only a small number
of images in F450W, F606W and F814W, which are background-limited, were
affected by the Earth cross.  On the other hand, the majority of the
F300W images show a visible Earth cross. 

Images affected by the Earth cross in F450W, F606W, and F814W were
corrected by applying a median filter (width 21 pixels) to the image
obtained after subtracting the "expected" image produced by
{\task{blot}} as part of the cosmic ray rejection
(Section~\ref{sec:crrej}).  The median filter preserves the edges and
produces an adequate subtraction for a sharply varying background, such
as that produced by either a bias jump or the scattered Earth light. 
Direct inspection confirmed the subtraction to be satisfactory. 

The large number of F300W images affected by the Earth cross precluded
the use of the median-filtering procedure.  Instead, we obtained and
subtracted an ``average'' Earth cross with the same technique used to
correct for the variations in the sky brightness.  While not
perfect---since the properties of the Earth cross vary depending on the
orbital phase and pointing of HST---this technique did remove most
visible traces of the Earth cross, and produced a nearly flat final
image. 

\subsubsection {Other problems}

Bias jumps are small, sharp variations in the bias voltage that produce
a distinctive change of background (usually by less than 0.5 DN) on a
row boundary.  Bias jumps can be corrected using the same technique of
subtracting a median box-filtered residual used for the Earth cross in
F450W, F606W, and F814W.  Cosmic rays were identified properly in the
vast majority of cases.  A handful of exceptions---usually due to a very
large cosmic ray event---were identified and masked by hand.  Similarly,
moving targets usually were flagged properly by the cosmic ray rejection
routine, but in some cases they left visible residuals.  These were also
masked by hand.  This special processing allowed us to recover nearly
all of the images with anomalies, and only seven of the 260 images were
completely excluded from the final combination. 

\subsection {Final Combination}

The parameters chosen for the final drizzling were similar to those for
the HDF-N image combination wherever possible.  We used a pixel scale of
0.4 WF pixels, and a footprint of 0.5 WF pixels for the Wide Field
cameras and of 0.8 PC pixels for the Planetary Camera.  We produced a
noise image appropriate for large-area scaling, as described below; the
Appendix describes how the noise can be predicted for small areas. 

\subsubsection {Weight and noise maps}\label{sec:definenoise}

Optimal image combination requires weighting each input exposure in
inverse proportion to the square of the noise per pixel.  The
{\task{drizzle}} task accepts a weight image for each input exposure,
and produces an output weight image which can be used to compute the
noise in the final combined image.  In practice, we used weights that
reflect only the noise due to the {\it background}, both sky and
detector, without considering the local signal due to individual
objects.  While it is possible in principle to use schemes that obtain
an optimal combination while taking into account the presence of local
signal, their implementation is very complex when images are shifted at
the subpixel level.  The ``natural'' solution of using the measured
signal level to estimate the noise in each pixel can produce biased
results.  The background noise is the appropriate quantity to consider
when estimating the significance of detection, which implies rejection
of the null hypothesis that there is no object.  Optimal combination
based on the background noise also provides the best sensitivity for
faint objects, for which the gain from an optimal combination is most
important.  Combination based on background noise is near-optimal for
{\it all} objects, regardless of their brightness, in
background-dominated images such as F606W and F814W, and for all but the
brightest objects in images where the read noise is a significant
component of the noise. 

Because the output pixels in {\task{drizzle}}d images are not
independent, the ``noise'' in individual pixels does not have the same
meaning as for the input images, and noise does not scale as it would
for individual pixels.  In particular, when estimating the noise over an
area, the correlation between pixels causes the total noise to be larger
than the quadratic combination of the noise in individual pixels. 

In order to estimate the significance of detection and the quality of
the photometry of individual objects, the quantity of interest is the
rms noise expected in the total (background) signal over the typical
area covered by an object.  To provide a natural way of computing this
quantity, we have designed the weight maps for {\task{drizzle}} so that
they can be readily used to estimate the noise over a sufficiently large
area.  It is useful to define the {\it equivalent single-pixel noise} $
\bar \sigma_i $, which is the single-pixel noise that would be expected
for uncorrelated pixels with the same total signal.  The quantity $ \bar
\sigma_i $ can be determined from the weight $ W_i $ in pixel $ i $, as
given in the weight image:
 \begin{equation}
 \bar \sigma_i = 1 / \sqrt {W_i} 
 \end{equation}

On large scales, the expected rms variation of the total signal is
simply the sum in quadrature of the equivalent single pixel noise. 
Thus, the variance of the total signal over a sufficiently large area $
A $ is:
 \begin{equation}
 \sigma^2_A = \sum_{i \in A} {\bar \sigma}^2_i
 \end{equation}

Note that the variance in the single-pixel signal is substantially {\it
smaller} than $ \bar\sigma $.  On the other hand, the single-pixel
variance is artificially small due to the way the single-pixel signal is
constructed.  The Appendix gives the expected scaling and relations
between single-pixel noise and noise over larger areas. 

\subsubsection {Definition of Weight for Input Images}

The weight of each input pixel is defined as follows:
 \begin{equation}
 {\rm Var} = [f (D+B) /\gamma + \sigma_{\rm read}^2] / (f^2 \, t^2)
 \end{equation}
 \begin{equation}
 W = 1 / ({\rm Var}  \times  {\rm Scale}^4) 
 \end{equation}
where $ \rm Scale $ is the linear size of the output pixel in units of
the input pixel ($ 0.4 $ for the WF cameras, $ 0.875 $ for the PC); $ D
$ and $ B $ are the total signal in dark current and background,
respectively, averaged over the image and expressed in counts (DN) per
pixel; $ \gamma $ is the gain in $ e^-/{\rm DN} $ (about 7); $
\sigma_{\rm read} $ is the read noise, also in DN/pixel; $ f $ is the
{\it inverse} flat field, the quantity that appears in the WFPC2 ``flat
field'' reference files; and $ t $ is the exposure time in seconds.  The
images were scaled by the exposure time in order to obtain a final image
in counts per second. 

\subsection {Final drizzling}

The final image was obtained by running the task {\task{drizzle}} with
parameters PIXFRAC of 0.5 for the WF images, and 0.8 for the PC, and
SCALE of 0.4 for the WF2, resulting in a linear scale of $ 0\farcs 03986
$/pixel.  The parameter SCALE for the other cameras was chosen to
achieve the same linear pixel scale as for the WF2.  The linear scale of
the final image was verified {\it a posteriori} using the astrometric
reference stars provided by the US Naval Observatory; see
Section~\ref{sec:astrometry}.  The size of the final image is 4096 x
4600 pixels, or $ 163'' \times 183'' $---enough to include all the area
with valid data.  The intensity is expressed in units of counts per
second, and the weight images defined above are used to obtain a final
weight image that could be used to determine the noise per unit area. 

\section {Quality Control of the Combined Images}

\subsection {Noise Properties}

Because neighboring pixels are strongly correlated, the noise in the
HDF-S images is not well described by its single-pixel variance.  On
large scales, the expected rms variation of the total signal scales with
the linear size $ N $ of the region, and in the limit of very large
area, it approaches $ N \bar \sigma $, where $ \bar\sigma $ is the
equivalent single-pixel noise defined in Section~\ref{sec:definenoise}. 
Due to the noise correlation, the rms variation expected for the signal
on smaller scales is substantially smaller.  The calculations reported
in the Appendix (Equation~\ref{eq:generalnoisescaling}) predict the
approximate behavior of the total noise on linear scale N as:

 \begin{equation} \label{eq:noisescalewithN}
  \sigma (N) = \bar\sigma \times \cases {44/75 & $N=1$ \cr N \times [1 - 5/(12\, N)] & $ N > 1 $ \cr}
 \end{equation}
 
The actual noise has been measured by computing the rms variation
in the total signal in several apparently empty areas of each image. 
The results are shown in Figure~\ref{fig:noise}, and scale similarly to
the prediction of Equation~\ref{eq:noisescalewithN}.  The vertical line
in the Figure indicates the linear size corresponding to 0.2 square
arcseconds, the box for which limiting magnitudes are given in
Table~\ref{table:limitingmag}. 

On small scales, the measured noise agrees extremely well with the
predictions, indicating the high quality of the data and of the noise
model.  On larger scales, the measured noise grows somewhat faster than
the model prediction, and is up to 20\% larger than predicted on a
linear scale of 15 pixels.  This discrepancy is most likely due to a
combination of very faint sources and small irregularities in the background.

\subsection {Quality of the Point Spread Function}

Figure~\ref{fig:psf} shows the radial profile of the final Point Spread
Function (PSF) of well-exposed objects of very small size (most likely a
star) in each of the four filters in the region covered by the Wide
Field Cameras.  The full-width at half-maximum of the best-fitting
Gaussian is about $ 0\farcs13 $ ($ 0\farcs16 $ in F300W), in good
agreement with the PSF size in individual images.  The PSF is
significantly smaller ($0\farcs088$) in the region covered by the
Planetary Camera.  The smoothness and tightness of the PSF demonstrates
the quality of the image combination, and especially of the relative
pointings we determined.  The size of the PSF is a crucial component of
the star-galaxy separation. 

\subsection {Background Variations}

At the depth of the final HDF-S images, very small irregularities in the
background can become visible.  Known sources of instrumental background
variations include imperfections in the flat fields, non-uniform
illumination, especially the Earth cross, and the so-called dark glow
(cf.~WFPC2 Instrument Handbook).  Slight fluctuations in the effective
background can be seen with heavy smoothing of the final images. 
However, thanks to the additional background correction steps taken in
our analysis, the first two sources of irregularities have been reduced
substantially compared to earlier combinations, and the residual
fluctuations are extremely small. 

The total effective backgrounds and their measured fluctuations are
given in Table~\ref{table:background}.  For F300W, the background
variations are at the 1.5\% level, and are dominated by residual
scattered Earth light, which could not be removed entirely.  In the
other filters the background variation is much smaller, about
0.2\%--0.6\%, and appears consistent with a 2--3\% curvature in the dark
current subtraction which could be related to the residual dark glow. 

\subsection {Astrometric Calibration}\label{sec:astrometry}

The final image was astrometrically calibrated using four stars with
absolute astrometry determined by the Naval Observatory in the Hipparcos
reference frame (Gauss {\etal} 1996, Zacharias 1997).  The stars are
listed in Table~\ref{table:astrometry}, with both their astrometric and
pixel positions.  After fitting for shifts, scale, and rotation, the rms
residual is 15 mas in RA and 5 mas in Dec.  The measured pixel size is $
0\farcs 03986$.  The World Coordinate System (WCS) parameters in the
header were updated to reflect the best-fitting solution.  The main
uncertainty in the absolute astrometry is in the systematics of the
positions of the reference stars, estimated to be less than 40 mas. 

\subsection {Impact of Charge Transfer Efficiency}

After several years in orbit, the Charge Transfer Efficiency (CTE) of
WFPC2 has decreased measurably (Whitmore {\etal} 1999), probably as a
result of the cumulative effect of particle hits on the detectors. 
Consequently, the effective photometric response of the detector
decreases at higher row and column numbers.  The impact of the CTE
decrease depends on source position, brightness, and shape, as well as
on the background.  Whitmore {\etal} (1999) offer an empirical assessment
of the photometric impact of CTE for point sources, as well as an
approximate correction.  A similar program is currently underway for
extended sources, but no results are available yet. 

If extended (but small) sources behave similarly to point sources, CTE
corrections for HDF-S sources are expected to be modest but not
insignificant, in the neighborhood of 10--20\%.  The prescriptions of
Whitmore et al (1999) predict a median point-source CTE correction of
18.7\% in F300W, 13.0\% in F450W, 9.2\% in F606W, and 10.4\% in F814W
(Figure~\ref{fig:whitmore_cte}).  These should be regarded as upper
limits, since extended sources are likely to be affected by loss due to
CTE less than point sources. 

An empirical indication of CTE-like effects is in the color of sources. 
The color distribution of galaxies is expected to be intrinsically
independent of location in the field of view, except for possible
selection effects related to the varying depth of observations.  If CTE
losses do in fact affect the HDF-S data, sources that typically lie
near the top of each chip should appear fainter.  The loss of signal is
expected to be relatively small in F606W, by virtue of its high
background and relatively large count rate.  On the other hand, the loss
of signal should be more pronounced in the blue filters---especially
F300W---which have a low natural background and fewer counts per source. 
Thus, one might expect the color distribution to be biased towards
redder colors at high $ x $ and $ y $ whenever F300W is involved. 

No clear evidence for such a trend exists in the data.  In
Figure~\ref{fig:umb_cte} we plot the measured $ U - B $ color against
the approximate coordinates in the original images.  Full circles (solid
lines) correspond to objects brighter than $ U = 26.0 $, and crosses
(dashed lines) to objects in the range $ 26.0 < U < 27.5 $; the lines
represent straight linear regressions to the points shown.  Any changes
in mean color with position are relatively small, and in opposite
directions in the two cases: galaxies appear somewhat redder at high $ x
$, and very slightly {\it bluer} at high $ y $, which is contrary to
expectations.  In both cases, the regression for fainter objects ($ 26 <
U < 27.5 $) is tilted slightly more negatively than that for brighter objects
($ U < 26 $), again contrary to expectations--fainter sources should be
more sensitive to CTE, and thus appear comparably redder at high $ x $ and $
y $.  Similar results are seen in other colors as well.  The mean color
varies with position differently in each detector, suggesting that the
origin of this variation is not likely to be instrumental, but rather
could reflect correlated variations in the intrinsic color distribution
of the sources on the sky.  We conclude that the color distribution of
galaxies is too noisy for an independent determination of the effects of
CTE, but CTE is unlikely to affect typical galaxy colors by more than
0.2 magnitudes. 

\section {Object Identification and Measurement}

The version 2 catalogs were produced using the Source Extractor
({\sextractor}; Bertin \& Arnouts 1996) package, version 2.1.0.  The HDF-N
catalog of Williams {\etal} (1996) was constructed using FOCAS.  To
allow direct comparison of the two data sets we have created a new
catalog of the HDF-N data set using {\sextractor} with the same parameters
used for HDF-S.  We shall limit our discussion here to the version 2
catalogs, which supersede the initial catalog released on the World Wide
Web in December 1998.  The differences between the two catalogs are
relatively minor, the most important being that version 2 images use
more of the data with a better removal of the instrumental signatures. 

A major motivation for using {\sextractor} was its incorporation of
weight maps and rms maps in modulating the source detection thresholds
and in determining the photometric errors.  While this is in principle
straightforward, there are some subtleties in dealing with the data for
both Hubble Deep Fields (HDF) that we will note below. 

Source detection and deblending were carried out on the
inverse-variance-weighted sum of the F606W and F814W drizzled images. 
While other techniques for weighting the photometric information from
the different bands are possible (e.g., Szalay {\etal} 1999), this
summed image provides the maximum limiting depth for most sources.  The
resulting source positions and isophotes are then used for subsequent
photometric analysis in each of the individual bands.  The combined
F606W+F814W image is significantly deeper than any of the individual
images, so few normal objects should be missed by using it to define the
object catalog for all bands.  It is possible, however, that objects
with very strong emission lines in the F450W or F300W bands could have
escaped detection. 

Source detection with {\sextractor} follows a standard ``connected pixel''
algorithm used by most such programs.  To identify sources, the
detection image is first convolved with a fixed smoothing kernel, for
which we use a circular Gaussian with FWHM = 3.6 pixels.  Pixels with
convolved values higher than a fixed threshold are marked as potential
sources.  This threshold is in units of $\sigma_{\rm sky}$, where
$\sigma_{\rm sky}$ as a function of position comes from the input RMS
map.  The variation of S/N as a function of position is thus
automatically taken into account. 

After thresholding, regions consisting of more than a certain number of
contiguous pixels (including diagonals) are counted as sources.  For the
HDF data, the source detection threshold was set to 0.65 (per pixel in
the convolved image), and the minimum area to 16 drizzled pixels, or
0.025 square arcseconds.  The sources are then examined for
sub-components using the segmentation scheme described by Bertin \&
Arnouts (1996).  Each source is reconsidered at a series of 32
logarithmically stepped detection thresholds between the original
detection threshold and saturation.  If the source
divides into two sources with a flux ratio greater than 0.03 at one of
the thresholds, the source is considered as two separate objects.  This
splitting continues for each of successively higher thresholds. 

For the HDF-N FOCAS catalog, Williams {\etal}~(1996) listed all of the
parents and daughters of this kind of splitting process.  In the
subsequent analysis, such as computing galaxy counts and finding
candidate Lyman-break galaxies, color was used as an additional
criterion to determine whether to combine two neighboring objects or
keep them separate.  {\sextractor} does not provide the option of
outputting the whole splitting hierarchy.  The catalog here thus offers
a bit less flexibility for post-processing than the original HDF-N
catalog (Williams {\etal}~1996). 

After extensive tests of {\sextractor} with different input parameters, it
became clear that no single set of parameters could satisfy the desire
to create a catalog that would reach the faintest reasonable depth for
isolated objects, but which at the same time did not include spurious sources
around stellar diffraction spikes and did not merge faint companions of
bright galaxies with the bright galaxy itself.  The detection parameters
described above were acceptable for most of the image, with only a few
cases of clearly spurious sources or inappropriate merging. 
Nevertheless, because the HDF catalogs will be used to find rare objects
and to identify candidates for spectroscopy, it seemed appropriate to
take some care to fix these problems.  The simplest solution was to run
{\sextractor} a second time with a higher detection threshold and to
selectively remove objects from the first catalog and add them in from
the second catalog.  The detection threshold of the second run was five
times higher than for the first run.  Objects that were clearly spurious
(which were all on top of diffraction spikes) were removed from the
catalogs.  Objects that were over-merged in the first catalog were
replaced with their corresponding split entries from the second catalog. 
The end result is that magnitudes and other photometric quantities for
objects from the second {\sextractor} run refer to a brighter isophote than
from the first run.  While this is not ideal, it is better than the
alternative of miscataloging clearly distinct objects as part of the
same entity.  A total of 25 sources (out of 2650) were removed from the
first run of the WFPC-2 catalog, and 32 sources were added in from the
second run.  Objects from the second {\sextractor} run have catalog
numbers greater than 10000. 

It is worth commenting on the use of the HDF error map in constructing
the catalog. This affects both the detection and the photometry
of sources. For the case of a single, cosmetically clean CCD image, the 
uncertainty of a galaxy flux is 
 \begin{equation}
		(C + B n + \sigma_{\rm read}^2 n)^{1/2},
 \end{equation}
 where $n$ is the number of pixels in the galaxy image, $C$ is the total
number of electrons detected from the galaxy, $B$ is the estimated
background (sky plus dark current) underlying the galaxy in $e^-$/pix,
and $\sigma_{\rm read}$ is the rms read noise in $e^-$/pix. 
In this approximation, we ignore the uncertainties in the local background
estimate. 

For the HDF, the effective exposure time varies from pixel to pixel, and
it is useful to use a noise model to compute the effective contributions
from background and read noise in each pixel.  As described in
Section~\ref{sec:definenoise}, the weight map for each image was
constructed taking into account the noise sources at the mean sky level
in each pixel, the variations in exposure time and detector sensitivity,
and the correlations between neighboring pixels introduced by the
resampling onto the final pixel grid.  The RMS map given to
{\sextractor} was $\sqrt{1/W}$, where $W$ is the weight map described in
Section~\ref{sec:definenoise}. 

At the source detection phase, the RMS map is used to modulate
the detection threshold. The detection threshold in counts 
for a given pixel is 
 \begin{equation}
	t_i = T  \sqrt{\sigma_i^2}
 \end{equation}
 where $T$ is the default threshold in units of $\sigma$, and
$\sigma^2_i$ is the variance at pixel $i$ computed from the RMS map. 
The variance map is convolved with the same kernel as the image as part
of the detection process. 

In computing photometric errors, the uncertainties for individual pixels
used in the sky estimate are simply set to the $\sqrt{\sigma_i^2}$
values in the variance map (in this case not convolved with the
detection kernel).  The variance map is also used in computing the total
number of electrons detected from the source.  The counts in the image
are multipled by gain and by the ratio of the variance in each pixel to
the median variance over the whole image.  This is the appropriate model
when the variations in S/N across the field are primarily due to changes
in exposure time, as is the case for the HDF.  Table~\ref{table:limitingmag}
gives the typical 10-$\sigma$ limiting magnitude in the deepest part of 
the Wide Field Camera images for an object of area $ 0.2 $ square arcsec.

\section {Catalog Parameters}

The catalog is presented in Table~\ref{table:catalog}, which contains a
subset of the photometry; extended catalogs with full photometry
information are available on the World Wide Web.  For each object we
report the following parameters:

{\bf ID:} The {\sextractor} identification number.  The objects in
the list have been sorted by Right Ascension (first) and Declination
(second), and thus are no longer in catalog order.  In addition, the
numbers are no longer continuous, as some of the object identifications
from the first {\sextractor} run have been removed.  Objects from the
second {\sextractor} run have had 10000 added to their identification
numbers.  These identification numbers provide a cross-reference to the
segmentation maps. 

{\bf HDFS\_J22r$-$60d:} The minutes and seconds of Right Ascension, and
negative arcminutes and arcseconds of Declination.  To these must be
added 22 hours (RA) and $-60$ degrees (Dec).  The catalog name of each
object can be determined from its coordinates, and has the form
HDFS\_Jhhmmss.ss$-$ddppss.s, where hhmmss.ss are hours, minutes and
seconds of Right Ascension, and ddppss.s similarly represent degrees,
arcminutes and arcseconds of (the absolute value of) declination.  Thus,
the first object in the catalog is named HDFS\_J223333.69$-$603346.0, at
RA $\rm 22^h 33^m 33\fs 69$, Dec $ -60^\circ 33\arcmin 46\farcs 0$,
Equinox J2000. 

{\bf x, y:} The x and y pixel positions of the object in the version 2
images. 

{\bf $m_i$, $\sigma(m_i)$, $m_a$:} The isophotal magnitude in the F814W 
image ($m_i$), its uncertainty ($\sigma(m_i)$), and the ``mag\_auto'' ($m_a$) 
in the same image.  The magnitudes
are given in the AB system (Oke 1971), where $m = -2.5 \log f_{\nu} -
48.60$.  (Our preferred notation for the magnitudes on the WFPC-2 AB
photometric system is $U_{300}, B_{450}, V_{606}$ and $I_{814}$ to avoid
confusion with the Johnson and Str\o mgren systems.  However space in
the tables does not allow us to use this convention here).  The
isophotal magnitude is determined from the sum of the counts
within the detection isophote, set to be 0.65$\sigma$.  The
``mag\_auto'' is an elliptical (Kron 1980) magnitude, determined from
the sum of the counts in an elliptical aperture.  The semi-major axis of
this aperture is defined by 2.5 times the first moments of the flux
distribution within an ellipse roughly twice the isophotal radius,
with a minimum semi-major axis of 3.5 pixels.

{\bf u-b:} Isophotal color, $ U_{300} - B_{450} $, in the AB magnitude
system, as determined in the detection-image isophote.  {\sextractor}
was run in two-image mode to determine the photometry in each separate
band image, using the weighted average of the F606W and F814W images as
the detection image.  For this and other colors, when the measured flux
in the bluer band is less than $2\sigma$, we list the 2-sigma
upper-limit to the color determined from the photometric errors in the
aperture.  If both bands are upper limits, no color is listed. 

{\bf b-v:} Isophotal color, $ B_{450} - V_{606} $, in the AB magnitude
system, as determined in the detection-image isophote. 

{\bf v-i:} Isophotal color, $ V_{606} - I_{814} $, in the AB magnitude
system, as determined in the detection-image isophote. 

{\bf $r_h$:} The half-light radius of the object in the detection image,
given in mas.  The half-light radius is defined by {\sextractor} as the
radius at which a circular aperture contains half of the flux in the
``mag\_auto'' elliptical aperture. 

{\bf s/g:} A star-galaxy classification parameter determined by a neural
network within {\sextractor}, and based upon the morphology of the
object in the detection images (see Bertin \& Arnouts 1996 for a
detailed description of the neural network).  Classifications near 1.0
are more like a point source, while classifications near 0.0 are more
extended. 

{\bf flags:} Flags are explained in the table notes, and include both
the flags returned by {\sextractor} and additional flags we added
while constructing the catalog.  Note that the overlap flag (a) is
set if more than 10\% of the area included in the mag\_auto calculation 
overlaps the isophotal area of a detected neighbor.
\section {Number Counts and Color-Color Plots}

The counts of galaxies as a function of apparent magnitude reflect both
cosmological curvature and galaxy evolution.  Galaxy counts are thus an
essential tool for testing cosmological models and the HDF data provide
the best statistics at faint magnitudes.  Figure~\ref{fig:countsonly}
shows a compilation of the galaxy counts from HDF-N, HDF-S, and other
surveys.  The HDF counts in this figure include no corrections for
incompleteness or biases in magnitude measurements.  Such corrections
are necessarily model-dependent, and are best included as part of model
comparisons to the data.  It is important to realize that these
corrections are likely to be substantial for any reasonable model.  The
downturn in the counts at faint $I$ magnitudes, for example, is probably
not significant given the uncertainties in the galaxy detection
efficiency near the limiting depth of the HDF images. 

Overall, the counts in the HDF-N and the HDF-S agree reasonably well. 
Figure~\ref{fig:ratioall} shows the ratio of the HDF-S counts to the
HDF-N counts in different bands down to approximately the 10$\sigma$
limiting magnitudes for typical galaxies.  The counts in the two fields
match well in the magnitude range shown.  At fainter magnitudes, the
HDF-S galaxy number densities are slightly lower than the HDF-N number
densities in all bands.  This is presumably due to the fact that the
HDF-S exposure times in the different bands were typically 84-90\% of
the HDF-N exposure times in all the bands and varied more across the
field; the deficit in the HDF-S at very faint magnitudes may be largely
due to incompleteness. 

Figures~\ref{fig:UdropplotN}--\ref{fig:BdropplotS} show color-color
diagrams similar to those used by Madau {\etal}  (1996) and others to
isolate high-redshift galaxies.  For this sample, we use only galaxies
with isophotal thresholds within 25\% of the mean threshold for the WF
chips.  This excludes galaxies found in the PC and a few galaxies along
the boundaries between the WF chips.  The total area used for finding
Lyman break galaxies is 4.24 arcmin$^2$.  We have also redone the
Lyman-break galaxy search on the new HDF-N catalog to ensure that the
different cataloging algorithms do not significantly affect the results. 
With the color selection used by Madau {\etal}, brighter than $B_{450} =
26.79$, the HDF-S has 74 U-dropout galaxies, likely to be at $2.2 < z <
3.5$, compared to 68 in the HDF-N.  With the slightly broader and deeper
color selection used by Dickinson (1998) and Steidel {\etal}  (1999)
there are 101 U-dropouts in the HDF-S and 106 in the {\sextractor} catalog
of HDF-N brighter than $R_{606+814} = 27$ In the $3.5 < z < 4.5$
B-dropout range, the HDF-S has 18 objects compared to 11 in the HDF-N
{\sextractor} catalog. 

The luminosity densities, and hence derived star-formation rates, also
differ between the HDF-N and the HDF-S.  Figure~\ref{fig:sfr_z1} shows
the star-formation rate (SFR) as a function of redshift derived from the
HDF-N and HDF-S and various ground-based surveys.  A variety of methods
have been used to determine the star-formation rate at $ z < 1 $, and a
complete review is beyond the scope of this paper; here we include
primarily those that are directly comparable with the high-redshift
results.  The star-formation rates are derived following the
prescription of Steidel {\etal} (1999).  We adopt a conversion
 \begin{equation}
   L_{1500} = 8.0 \times 10^{27} {{SFR} \over {M_\odot \rm yr^{-1}} {\rm \,erg\,s^{-1}\,Hz^{-1}}} 
 \end{equation}
 and apply a uniform extinction correction factor of 4.7 (Steidel
{\etal} 1999) to the measured UV luminosity densities.  To compute the
star-formation rate, we sum the fluxes of the detected objects, and
correct for undetected objects by adopting the luminosity-function
parameters of Steidel {\etal} (1999): $\alpha = -1.6$ and $m^*_{\cal R}
= 24.48$ at $z = 3.04$.  Adopting the cosmological parameters
$h,\Omega_m,\Omega_{\Lambda},\Omega_{\rm tot} = 0.65,0.3,0.7,1$ the
corresponding $m^*$ is $R_{606+814} = 24.195$ and $I_{814} = 24.96$ for
the U and B dropouts, respectively, where $R_{606+814}$ is the AB
magnitude in the average of the F606W and F814W images.  Using the
magnitude $m_{\rm AB,tot}$ computed from the total integrated flux of
the Lyman-break candidates, the star-formation rate density at a mean
redshift $z$ between two redshifts $z_1,z_2$ is given by
 \begin{equation}
   \log ({{SFR} \over {M_\odot \rm yr^{-1} Mpc^{-3}}} )  = 
		2 \log {{D_L(z)} \over {\rm cm}} - 0.4 m_{\rm AB,tot }
		- \log {{\Delta V(z_1,z_2)} \over {\rm Mpc^3}} - \log A
		+ \log \delta L - \log(1+z)
		- 34.516,
 \end{equation}
 where $D_L$ is the luminosity distance, $\Delta V$ is the total volume
between the two redshifts, $A$ is the area of the field of view in
arcsec, and $\delta L$ is the correction to the integrated luminosity
computed by integrating from the survey limit down to $0.1 L^*$. 

For both redshift ranges, the integrated star-formation rate density is
higher in the HDF-S than in the HDF-N.  At $z=4$ the SFRs for both
fields are significantly lower than that measured by Steidel {\etal}
(1999) from their ground-based survey.  This may indicate that the
assumption of a redshift independent luminosity function is faulty, or
it may indicate that strong clustering causes significant field-to-field
variations.  The SFR at $z \sim 4$ computed here from the HDF-N
{\sextractor} catalog is a factor of 2.6 lower than that shown in Fig.~9
of Steidel {\etal} (1999).  The differences are due in part to the 
different cosmology adopted here, and in part to the different
photometry in the {\sextractor} and FOCAS catalogs.  In any event, the
integrated SFR at $z \sim 4$ remains uncertain from these small fields,
and from the limited magnitude range probed by the ground-based survey. 
This is an area where a large-area survey from the HST Advanced Camera
for Surveys would be particularly beneficial. 

\section {Conclusions}

The interpretation of any astronomical image requires distinguishing the
true celestial signal from the artifacts, non-uniformities, and
non-linearities of the optics and detector.  In many cases, fortunately,
the scientific results can be extracted from the data with only a
rudimentary understanding of the behavior of the detector and the noise
properties of the images it produces.  However, the HDF-S is intended to
be a data set of lasting archival value for a wide variety of purposes
(some undoubtedly still to be conceived).  Thus, it is important to have
a lasting archival record of how the images were produced, and a
detailed description of technical peculiarities of the final images. 

Sections 1-6 of this paper outlined the technical features of the WFPC2
portion of the HDF-S campaign.  Significant departures from the HDF-N
campaign include a different dithering scheme (dictated primarily by the
simultaneous STIS observations of QSO J2233-606), a different cosmic-ray
rejection technique, and a slightly different weighting scheme for
combining the final images.  Calibration observations show significant
degradation of the WFPC-2 charge-transfer efficiency over the three
years since the HDF-N campaign.  However, tests for the symptoms of CTE
in the photometry of galaxies at different positions on the chips reveal
no significant trends, and we conclude that CTE affects galaxy
photometry at a level less than 0.2 mag. 

In Section 7, we compared the overall numbers of galaxies, and the
number of Lyman-break galaxy candidates, in the HDF-N and HDF-S.  For
this purpose a new HDF-N catalog was constructed in a manner identical
to that described for HDF-S.  Overall number counts between the two
fields are in excellent agreement.  The HDF-S shows more candidate
high-redshift objects, and a higher overall UV luminosity density at
$z > 2$; however the luminosity densities derived here are still
significantly lower than the $z = 4$ luminosity density found in the
Steidel {\etal}  (1999) groundbased survey.  The discrepancy could be due
either to field-to-field variations, or to a steepening of the faint end
of the Lyman-break galaxy luminosity function at $z \sim 4$ relative to
that at $z \sim 2.7$. 

\acknowledgments

It is a pleasure to thank the U.S. Naval Observatory CCD Astrograph
team and CTIO for making possible the astrometric link of the HDF-S
positions to the ICRS.  We are grateful as well to our many colleagues
who shared their ground-based data with us in advance of publication,
thus allowing us to optimize our observing strategies.  This work was
supported by several grants from the Space Telescope Science Institute,
which is operated by the Association of Universities for Research in
Astronomy, Incorporated, under NASA contract NAS5-26555. 

\appendix

\section {Noise correlation in drizzled images}

\subsection {Goals and definitions}

Under normal circumstances, the noise properties of CCDs are well
described by the variance of the signal detected in each pixel.  Both
Poisson and read noise are with good approximation uncorrelated from
pixel to pixel, and simple noise models can be used to accurately
predict the image variance. 

However, combining images with partial pixel overlap results in more
complex noise properties, regardless of the method used.  The main
reason is that a ``pixel'' in the output image is no longer a separate
physical entity from all other pixels.  The signal apportioned to it
from each input image can be a fraction of the signal in a number of
different physical pixels, some of which may also contribute to the
signal assigned to neighboring output pixels.  The signal in nearby
output pixels can thus be correlated; a full description of their noise
properties requires not only a variance image, but also a covariance for
each pair of correlated pixels. 

A correlation image can in principle be produced, but at a very high
cost in terms of computational intensity and complexity.  In practice,
rules describing the scaling of the effective noise as a function of the
area considered may often suffice.  The goal of this Appendix is to
illustrate some approximate rules that can be used to estimate the
effective noise in images produced by the {\task{drizzle}} method.  For
more details on the method see Fruchter and Hook (1997); a more general
derivation of the noise properties of {\task{drizzle}}d images is
presented in Fruchter and Hook (2000). 

In general, we will be interested in estimating the total noise within a
given output area, be it a single output pixel or the area covered by an
object of astronomical interest.  Since {\task{drizzle}} is a linear
method, the total signal in the area of interest is a linear combination
of the signal in input pixels from different images and with varying
weights; its variance can in principle be estimated by the appropriate
weighted combination of the variance of the input pixels.  Some of these
pixels will be included in their entirety, others---near the border of
the area of interest---only fractionally.  The covariance between two
areas is caused by pixels that are fractionally included in both, and
thus it is a perimeter effect.  If the area is sufficiently
large---enclosing a significant number of input pixels---the perimeter
effect can be neglected, and the variance scales with the area, as
expected for uncorrelated noise. 

An analytical approximation for the scaling of the noise properties in 
an area with its size can be obtained under a number of simplifying
assumptions.  We assume that all input images have similar properties
and have been combined with the same parameters; we also assume that the
input images have random pixel phases, so we can average over the
fractional position of the input pixel with respect to the output pixel. 
We also neglect rotations and corrections for geometric distortions,
assuming that the input pixels are all square and have the same size. 
Finally, the output area is assumed to consist of a square with an
integral number of output pixels to each side.  We expect that the above
assumptions, while not perfectly realized in our data, are sufficiently
close to reality that our results are applicable in practice. 

The parameters in our calculation are: the {\task{drizzle}} parameters
SCALE $ s $, or the size of each output pixel, and PIXFRAC $ p $, or the
footprint that {\task{drizzle}} assigns to each input pixel; and the
linear size $ l $ of the square output area.  All are expressed in units
of the size of the input pixel. 

We use $ x_{ij} $ to refer to the signal in pixel $ i $ on the $ j $-th
input image, and $ X_k $ to refer to the signal in pixel $ k $ in the
output image.  We further assume that the input images only contain a
flat background with a nominal flat-fielded signal of $ x $ per pixel,
and noise $ \sigma_{ij} $.  Since the input pixels are assumed
uncorrelated, their mean and covariance is:

\begin{eqnarray*}
 \avg{ x_{ij} } & = & x \\
 \avg{ x_{ij} x_{i'j'} } - x^2 & = & \delta_{ii'} \delta_{jj'}
     \sigma_{ij}^2
 \end{eqnarray*}

\subsection {Drizzled images}

The signal of each pixel in the output image is computed as
 \begin{equation}
 X_k = s^2 \cdot \sum_{ij} a_{ij,k} x_{ij} w_{ij} {\Big {/}}
   \sum_{ij} a_{ij,k} w_{ij}
            \, , 
 \end{equation}
 where $s^2 $ is the area of the output pixel in units of the input pixel,
and $ a_{ij,k} $ is defined as the fraction of the signal of
pixel $ i $ in image $ j $ that is apportioned to pixel $ k $ of the output
image.  By construction,
 \begin{equation}
 \sum_k a_{ij,k} = 1  \, . \label {eq:fluxconserv} 
 \end{equation}

The weight $ w_{ij} $ can be chosen arbitrarily, but we will in general
assume that it reflects the expected noise in the background and is flat
over small scales, i.e., $ w_{ij} = 1/\sigma_{ij} $ and $ w_{ij} $ is
nearly independent of $ i $ over small scales, except for rejected
pixels. 

The variance of $ X_k $ is not obtained simply by scaling the noise from
the input to the output pixels.  There are two reasons for this: first,
each input pixel is scaled by $ s^2 $, which reduces the variance in the
output pixel accordingly; second, for $ p > 0 $, each pixel may receive
partial contributions from multiple input pixels, even though the
individual input pixels may be larger than the individual output pixels. 
Therefore the $ X_k $ are {\it not} uncorrelated, and their individual
variance can be substantially lower than would be expected from a simple
area extrapolation; conversely, their variance increases with area
faster than the area itself.  Thus, while it is possible to define a
variance image for the $ X_k $, i.e., an image that contains the
expected variance of the signal for each pixel, this variance image
would {\it not} have the normal properties expected from a variance
image; namely, that the variance of the total signal over a number of
pixels is the sum of the variances of the individual pixels. 

Formally, we note that the $ X_k $ would be uncorrelated if only the
values 0 and 1 were allowed for the coefficients $ a_{ij,k} $, i.e., if
each input pixel were either included or excluded in its entirety.  The
correlation derives from the fact that the input pixels can be split
between different output pixels, so that the $ a_{ij,k} $ can have
values different from 0 and 1, and $ \sum_k (a_{ij,k})^2 < 1 $. 

On the other hand, most astrophysically useful noise measurements are on
scales much larger than one output pixel.  On large scales, the variance
will indeed scale approximately with the area, with a correction for
correlation that is approximately inversely proportional to the linear
size of the region considered, as will be shown later.  

Under these conditions, it is desirable to define an ``equivalent
variance per pixel'' that gives the correct noise on large scales, and
an expression that indicates the correction to be applied as a function
of the area.  We now proceed to derive the proper weight definition to
achieve the correct noise estimate in the limit of infinite area
(Section~\ref{sec:weightinfinite}) and to show how the noise should be
corrected for finite areas (Sections~\ref{sec:variancereduction}
and~\ref{sec:noisescaling}). 

\subsection {Scaling of noise for large areas}\label{sec:weightinfinite}

The drizzle task provides a total weight per pixel that can be used to
assess the relative statistical weight of each output pixel,
 \begin{equation}
 W_k = \sum_{ij} a_{ij,k} w_{ij} \, .  
 \end{equation}
Let us assume for the moment that the $ W_k $ are the appropriate weights
for the signal in pixel $ k $.  Then the total weighted signal $ X_A $
in a large area $ A $ will be expressed as:
 \begin{equation}
 X_A = \sum_A (A/A_k) X_k W_k \Big / \sum_A W_k \, .  
 \end{equation}
 This can be re-expressed as:
 \begin{equation}
 X_A = \sum_{k \in A} (A/A_k) s^2 \sum_{ij} a_{ij,k} x_{ij} w_{ij}
    \sum_{k \in A} \sum_{ij} a_{ij,k} w_{ij} \, , 
 \end{equation}
 where we have implicitly assumed that $ A $ includes only full pixels of
the output image; otherwise the expression must be modified to include
partial pixels.  Sum over $ k $ and define
 \begin{equation}
 a_{ij,A} = \sum_{k \in A} a_{ij,k} \, ,  
 \end{equation}
 then,
 \begin{equation}
 X_A = (A/A_k) s^2 \sum_{ij} a_{ij,A} x_{ij} w_{ij} \Big /
                 \sum_{ij} a_{ij,A} w_{ij} \, , 
 \end{equation}
 and
 \begin{equation}
 \variance (X_A) = (A/A_k)^2 s^4 \sum_{ij} a_{ij,A}^2 w_{ij} \Big /
      \Bigl ( \sum_{ij} a_{ij,A} w_{ij} \Bigr )^2 \, , 
 \end{equation}
 where we have used the fact that $ \variance(x_{ij}) = 1/w_{ij} $, and
that the $ x_{ij} $ are uncorrelated. 

If the area $ A $ is sufficiently large, the majority of the pixels $ i
$ will either be entirely inside $ A $ or entirely outside; therefore $
a_{ij,A} $ is either 0 or 1 (cf.~Equation~\ref{eq:fluxconserv})
 for most $ i,j $.  Thus,
 \begin{equation}
 a_{ij,A}^2 \approx a_{ij,A} \, , 
 \end{equation}
 and, in the limit $ A \rightarrow \infty $,
 \begin{equation}
 \variance(X_A) \approx (A/A_k)^2 s^4 \Big / \Bigl (\sum_{ij} a_{ij,A}
        w_{ij}\Bigr)
     \approx (A/A_k)^2 s^4 \Big / \Bigl (\sum_{k \in A} W_k\Bigr ) 
         \, . 
 \end{equation}
 If one defines $ \hat W_k = W_k / s^4 $, and $ N_A = A/A_k $ is the
number of pixels included in area $ A $, we then have:
 \begin{equation}
 \variance (X_A) \approx N_A / \avg { \hat W_k }_A
     \label{eq:areavariance} \, , 
 \end{equation}
 which is precisely the scaling we expect for large area with uncorrelated
noise.  This formula can probably be generalized to the case in which input
images have different scales, by defining $ \hat W_k = \sum_{ij}
a_{ij,k} w_{ij} / s_j^4 $ ($ s_j $ being the scale of the $ j $-th input
image); however, we have not verified this formally.

\subsection {Area correction: the noise in a finite area}

The scaling formula above only works if the area $ A $ is very large. 
For smaller areas the variance can be predicted by a suitable
modification of Equation~\ref{eq:areavariance}.  It is convenient to
define the ``variance reduction factor'' $ F_A $ as:
 \begin{equation}
 F_A = \sum_{ij} a_{ij,A}^2 \Big / \sum_{ij} a_{ij,A} \, .  
 %
 \end{equation}
 Since $ a_{ij,A} \leq 1 $, then $ F_A \leq 1 $ as well, and $ F_A = 1 $
if, and only if, all $ a_{ij,A} $ are either 0 or 1, which means that
the area $ A $ contains only whole input pixels.  Note that we have also
assumed that $ A $ contains only whole {\it output} pixels.  The
quantity $ F_A $ measures of how many ``pixel pieces'' are contained
in the area $ A $.  The denominator is simply $ M_A $, the total number
of pixels, including fractions, that are included in area $ A $.  The $
a_{ij,A} $ depend on the pixel kernel used, which in turn is defined by
the value of the parameter $ p $, the size of the pixel footprint in
{\task{drizzle}}.  If $ p $ is very small, then $ a_{ij,A} $ will be
typically either 0 or 1, and $ F_A \approx 1 $ even for small areas. 

Assuming that $ w_{ij} $ remains constant for areas $ \sim A $, we can
express $ \variance (X_A) $ as
 \begin{equation}
 \variance (X_A) \approx (A/A_k)^2 s^4 F_A \sum_{ij} a_{ij,A} w_{ij}
         \Big / \Bigl ( \sum_{ij} a_{ij,A} w_{ij} \Bigr )^2
     \approx (A/A_k) F_A / \avg{\hat W_k} \, .  
 \end{equation}
 This shows that the variance per unit area is reduced by a factor $ F_A
$ with respect to the large-area scaling. 

The value of $ F_A $ will depend in general on the {\task{drizzle}}
parameters SCALE and PIXFRAC, and on the exact placement of the area $ A
$ onto the input grid.  For the latter reason, $ F_A $ can vary
significantly from place to place, even for equal-area regions. 
However, it is possible to estimate the typical value of $ F_A $
assuming random placement of the area $ A $ onto the input grid.  We now
proceed to work out the special case in which the input pixels are
square and parallel to the output pixels, and the area $ A $ is a square
containing an integral number of output pixels. 

\subsection {Variance reduction factor for random
phases}\label{sec:variancereduction}

An analytic expression for the variance reduction factor $ F_A $ can be
obtained in the special case in which 1) the input and output grids are
parallel, 2) the area $ A $ consists of a square region with an integral
number of pixels on each side, and 3) the boundaries of input and output
pixels are at random relative phases. 

Two distinct formulae apply depending on whether the linear size $ l $
of the area $ A $ is bigger or smaller than the pixfrac $ p $, which is
the size of the kernel used for each input pixel.  (Both $ l $ and $ p $
are expressed in input pixels.) Tedious but elementary calculations show that:
 \begin{equation}
 F_A = \cases {
      [1 - p/(3l)]^2 & \quad $ ( l > p ) $ \cr
 (l/p)^2 \cdot [1 - l/(3p)]^2 & \quad $ ( l < p ) \, .  $ \cr
 }   
 \end{equation}

Substituting in the formula for the variance, we obtain:
 \begin{equation}
 \variance (X_A) = (A/A_k) \cdot [l/\max(l,p)]^2 \cdot
   \{1-[\min(l,p)/(3 \max(l,p)]\}^2 / \avg{\hat W_k} \, .  
 \end{equation}

From this expression, it is possible to derive the predicted noise over
scales ranging from a single pixel to very large areas. 

\subsection {Scaling with size of the region
used}\label{sec:noisescaling}

With uncorrelated signal, the variance in the total signal over an area
scales linearly with the area considered.  In the case we are
considering, the signal is correlated (to some extent) on all scales,
and thus its expected variance does not scale exactly with area.  We can
use the above formulae to determine the scaling expected. 

Consider the case of two areas $ A_1 $ and $ A_2 $, of linear
size $ l_1 $ and $ l_2 $, both larger than the pixfrac $ p $.  The ratio
of the variances of the total signals $ X_1, X_2 $ is
 \begin{equation}
 \variance(X_1) / \variance (X_2) = (l_1)^2 [1 - p/(3 l_1)]^2 /
     \{(l_2)^2 [1 - p/(3 l_2)]^2\} \, , 
 \end{equation}
 and the ratio of the rms $ \sigma_1 $, $\sigma_2 $ will be
 \begin{equation}
 \sigma_1 / \sigma_2 = (l_1/l_2) \cdot [1 - p/(3l_1)]/[1-p/(3 l_2)]
     \, . 
 \end{equation}

The ``natural'' ratio (for uncorrelated noise) is the ratio of the
linear sizes $ l_1 / l_2 $, thus the actual noise ratio differs from the
uncorrelated value by the factor
 \begin{equation}
 [1 - p/(3l_1)]/[1-p/(3 l_2)] \, .  
 \end{equation}

For example, consider the HDF-S values $ p=0.5 $, $ s=0.4 $, and take $
l_1 = 10 $, $ l_2 = 2 $ (both expressed in input pixels).  Then the
ratio is $ (1 - 0.5/30) / (1 - 0.5/6) = 1.072 $; the noise in an area $
10 \times 10 $ input pixels ($ = 25 \times 25 $ output pixels) is 1.072
times the noise expected in an area $ 2 \times 2 $ input pixels ($ 5
\times 5 $ output pixels). 

The noise scaling from a single output pixel to a large area can be
obtained by taking the smaller size to be one output pixel, or $ l_2=s
$.  If the scale $ s $ is smaller than the pixfrac $ p $, as is often
the case, then
 \begin{equation}\label{eq:scalingnoise}
 \sigma_1 / \sigma_2 = (l_1/s) (p/s) [1 - p/(3 l_1)] / [1-s/(3p)]
 \, .  
 \end{equation}

If the size of the larger area in {\it output} pixels is $ m = l_1/s $, then
 \begin{equation}
 \sigma_1 / \sigma_2 = m (p/s) [1-p/(3ms)] / [1-s/(3p)]
   \, .  
 \end{equation}
 This differs from the ``area'' scaling by a factor $ (p/s) [1-p/(3ms)] /
[1-s/(3p)] $, which becomes $ (p/s) / [1-s/(3p)] $ as the larger area
grows to infinity.  For the HDF-S values, the large-area noise is larger
than the scaling from single-pixel by the factor $ (0.5/0.4) / (1 -
0.4/1.5) = 75/44 = 1.7045 $ in the region covered by the Wide Field cameras.
In the region covered by the Planetary Camera, $ p = 0.8 $ and $ s = 0.8755 > p $,
thus the large-area noise exceeds the value inferred from the single-pixel noise
by the factor $ 1 / [1 - p/(3s)] = 1.4380 $.

More generally, the scaling from single-pixel noise to infinity is given
by the factor
 \begin{equation}\label{eq:generalnoisescaling}
 \sqrt{F_A} = \cases {(s/p) [1 - s/(3p)] & $ (s < p) $ \cr
           [1 - p/ (3s)] & $ (p < s) \, .  $ \cr} 
 \end{equation}

Some special cases are
 \begin{equation}
 \sqrt {F_A} = \cases{
   5/6             & $ (p = 0.5 s) $ \cr
   2/3             & $ (p = s) $ \cr
  44/75 = 1/1.7045 & $ (p=0.5, s=0.4) $ [HDF-S, Wide Field cameras] \cr
  14/27 = 1/1.9286 & $ (p=0.6, s=0.4) $ [HDF-N, Wide Field cameras, Version 1] \cr
  0.6954 = 1/1.4380 & $ (p=0.8, s=0.8755) $ [HDF-S, Planetary Camera].  \cr}
 %
 \end{equation}

\clearpage

\begin {references}

\reference{} Abraham, R.~G., Tanvir, N.~R., Santiago, B.~X., Ellis, R.~S.,
Glazebrook, K., {van den Bergh}, S.~1996, MNRAS, 279, L47

\reference{} Bershady, M.~A., Lowenthal, J.~D., Koo, D.~C.~1998, ApJ, 505, 50

\reference{} Bertin, E., Arnouts, S.~1996, AApS, 117, 393

\reference{} Connolly, A.~J., Szalay, A.~S., Dickinson, M., SubbaRao, M.~U., Brunner
R.~J.~1997, ApJ, 486, 11

\reference{} Cowie, L.~L., Songaila, A., Barger, A.~J.~1999, AJ, 118, 603

\reference{} Dickinson, M.~1998, in {\it The Hubble Deep Field}, eds.~M.~Livio,
S.~M.~Fall, and, P.~Madau (Cambridge: Cambridge University Press), P.~219

\reference{} Djorgovski, S., Soifer, B.~T., Pahre, M.~A., Larkin, J.~E., Smith
J.~D., Neugebauer, G., Smail, I., Matthews, K., Hogg, D.~W., Blandford
R.~D., Cohen, J., Harrison, W., Nelson, J.~1995, ApJ, 438, L13

\reference{} Ferguson, H.~C., Baum, S.~A., Brown, T.~M., Busko, I., Carollo, M., et
al.~2000, in preparation

\reference{} Fruchter, A., Bergeron, L.~E., Dickinson, M., Ferguson, H.~C., Hook
R.~N., et al.~2000, in preparation

\reference{} Fruchter, A.~S., Hook, R.~N.~1997, in
{\it Applications of Digital Image Processing XX}, ed.~A.~Tescher,
Proc.~S.P.I.E. vol.~3164, 120

\reference{} Fruchter, A.~S., Hook, R.~N., Busko, I.~C., Mutchler, M.~1997, 
in {\it Proceedings of 1997 HST Calibration Workshop}, eds.~S.~Casertano et al.
(Baltimore: STScI), 518

\reference{} Fruchter, A.~S., Hook, R.~N.~2000, in preparation

\reference{} Gallego, J., Zamorano, J., Aragon-Salamanca, A., Rego, M.~1995, ApJ,
455, L1

\reference{} Gardner, J., Baum, S.~A., Brown, T.~M., Carollo, C.~M., Christensen, J.,
et al.~2000, AJ, 119, 486

\reference{} Gardner, J.~P, Sharples, R.~M., Carrasco, B.~E., Frenk, C.~S.~1996,
MNRAS, 282, L1

\reference{} Gardner, J.~P., Cowie, L.~L., Wainscoat, R.~J.~1993, ApJ, 415, L9

\reference{} Gauss, F.~S., Zacharias, N., Rafferty, T.~J., Germain, M.~E.,
Holdenried, E.~R., Pohlman, J.~W., Zacharias, M.~I.~1996, BAAS

\reference{} Gilmozzi, R., Ewald, S., Kinney, E., 1995, The Geometric
Distortion for the WFPC2 Cameras, WFPC2 Instrument Science Report 95-02
(Baltimore: STScI)

\reference{} Holtzman, J.~A., Hester, J.~J., Casertano, S., Trauger, J.~T.,
Ballester, G., Burrows, C.~J., Clarke, J.~T., Crisp, D., Gallagher
J.~S.~III, Griffiths, R.~E., Hoessel, J.~G., Mould, J.~R., Scowen, P.~A.,
Stapelfeldt, K.~R., Watson, A.~M., Westphal, J.~A.~1995, PASP, 107, 156

\reference{} Huang, J.-S., Cowie, L.~L., Gardner, J.~P., Hu, E.~M., Songaila, A.,
Wainscoat, R.~J.~1997, ApJ, 476, 12

\reference{} Kron, R., G. 1980, ApJS, 43, 305

\reference{} Lilly, S.~J., Cowie, L.~L., Gardner, J.~P.~1991, ApJ, 369, 79

\reference{} Lilly, S.~J., {Le F\'evre}, O., Hammer, F., Crampton, D.~1996, ApJ,
460, L1

\reference{} Lucas, R.~A., Baum, S.~A., Brown, T.~M., Casertano, S., {de Mello}, D., et
al.~2000, in preparation

\reference{} Madau, P., Ferguson, H.~C., Dickinson, M., Giavalisco, M., Steidel,
C.~C., Fruchter, A.~S.~1996, MNRAS, 283, 1388

\reference{} McLeod, B.~A., Rieke, M.~J.~1995, ApJ, 454, 611

\reference{} Minezaki, T., Kobayashi, Y., Yoshii, Y., Peterson, B.~A.~1998, ApJ,
494, 111

\reference{} Moustakas, L.~A., Davis, M., Graham, J.~R., Silk, J., Peterson, B.~A.,
Yoshii, Y.~1997, ApJ, 475, 455

\reference{} Oke, J.~B. 1971, ApJ, 170, 193

\reference{} Postman, M., Lauer, T.~R., Szapudi, I., Oegerle, W.~1998, ApJ, 506,
33

\reference{} Steidel, C.~C., Adelberger, K.~L., Giavalisco, M., Dickinson, M.,
Pettini, M.~1999, ApJ, 519, 1

\reference{} Sullivan, M., Treyer, M.~A., Ellis, R.~S., Bridges, T.~J., Milliard, B.,
Donas, J.~2000, MNRAS 312, 442

\reference{} Szalay, A.~S., Connolly, A.~J., Szokoly, G.~P. 1999, AJ, 117, 68

\reference{} Thompson, R.~I., Storrie-Lombardi, L.~J., Weymann, R.~J., Rieke, M.~J.,
Schneider, G., Stobie, E., Lytle, D.~1999, AJ, 117, 17

\reference{} Tresse, L., Maddox, S.~J.~1998, ApJ, 495, 691

\reference{} Voit, M., et al., 1997, HST Data Handbook, Version 3.0 (Baltimore: STScI)

\reference{} Whitmore, B., Heyer, I., Casertano, S.~1999, PASP, 111, 1559

\reference{} Williams, R.~E., et al.~2000, AJ, this issue

\reference{} Williams, R.~E., Blacker, B., Dickinson, M., Dixon, W.~V.~D., Ferguson
H.~C., Fruchter, A.~S., Giavalisco, M., Gilliland, R.~L., Heyer, I.,
Katsanis, R., Levay, Z., Lucas, R.~A., McElroy, D., Petro, L., Postman, M., Adorf
H.-M., Hook, R.~N.~1996, AJ, 112, 1335-1389

\reference{} Zacharias, N.~1997, AJ, 113, 1925

\end{references}

%% file: Casertano_realfigures.tex
 \begin{figure}[p]
 \centerline{\psfig{file=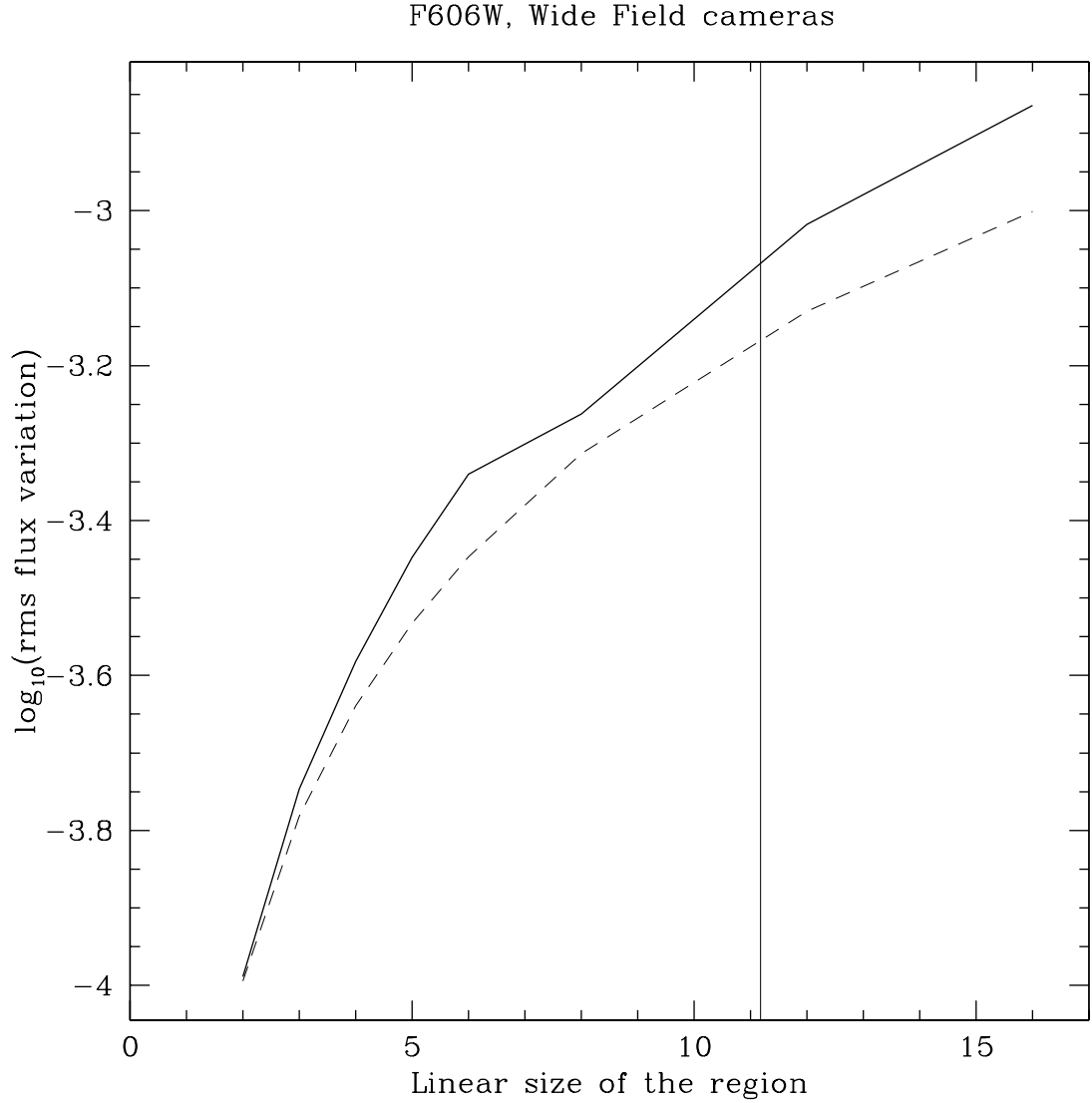,width=6in}}
 \figcaption[Casertano_fig1.ps] {Scaling of the measured rms noise as a
function of the linear size of the region (in output pixels).  The
measured values (solid) represent the median for the center of the three
Wide Field cameras in F606W.  The dashed line is the theoretical
prediction of Equation~\ref{eq:scalingnoise}, using the weight derived
in the final combination.  The deviation of the measured noise from the
prediction probably reflects correlated background
fluctuations.\label{fig:noise}}
 \end{figure}

 \begin{figure}[p]
 \centerline{\psfig{file=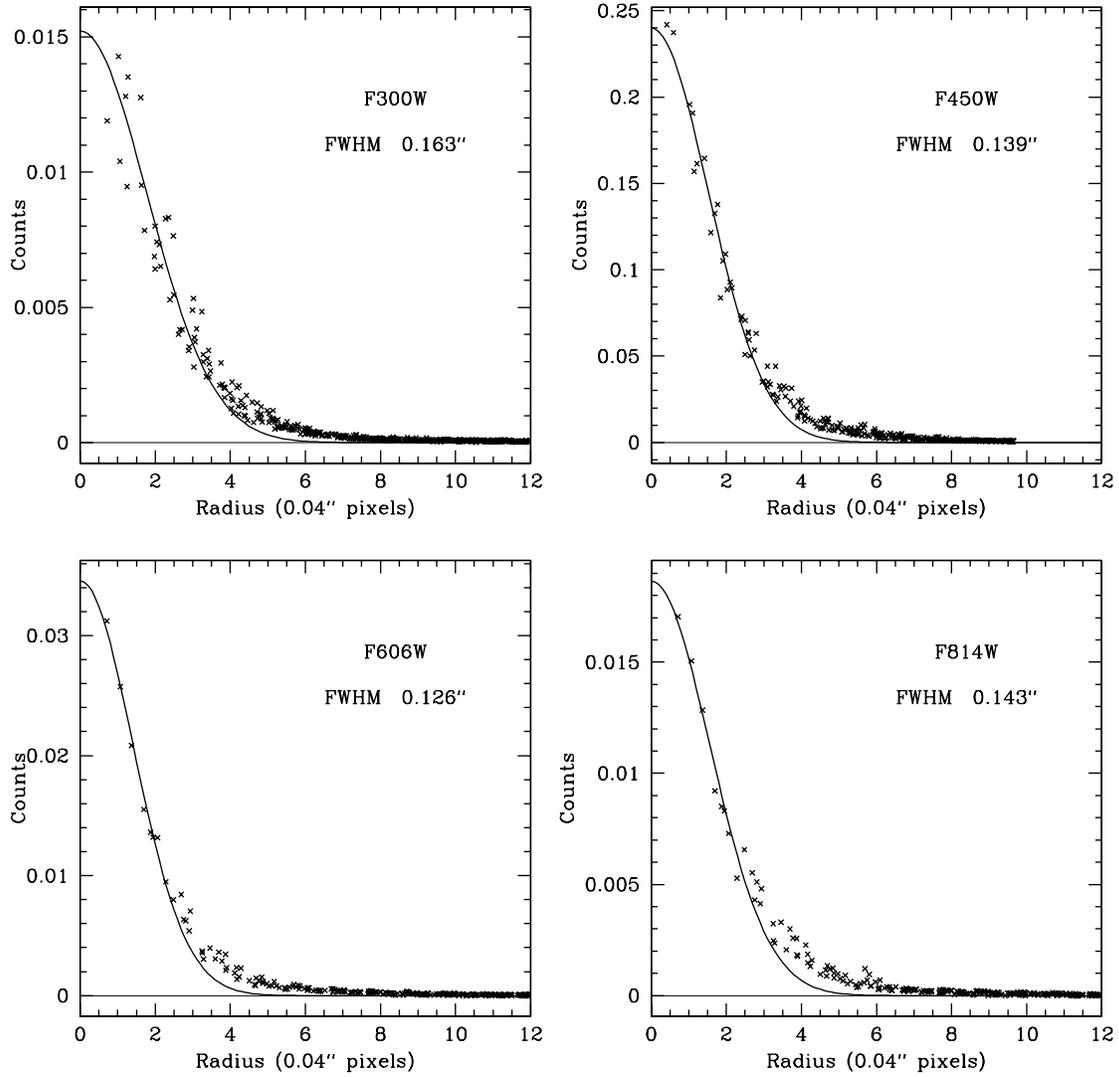,width=6in}}
 \figcaption[Casertano_fig2.ps] {PSF profile in the final image for
selected starlike objects in each of the four filters, in the region
covered by Wide Field cameras.\label{fig:psf}}
  \end{figure}

 \begin{figure}[p]
 \centerline{\psfig{file=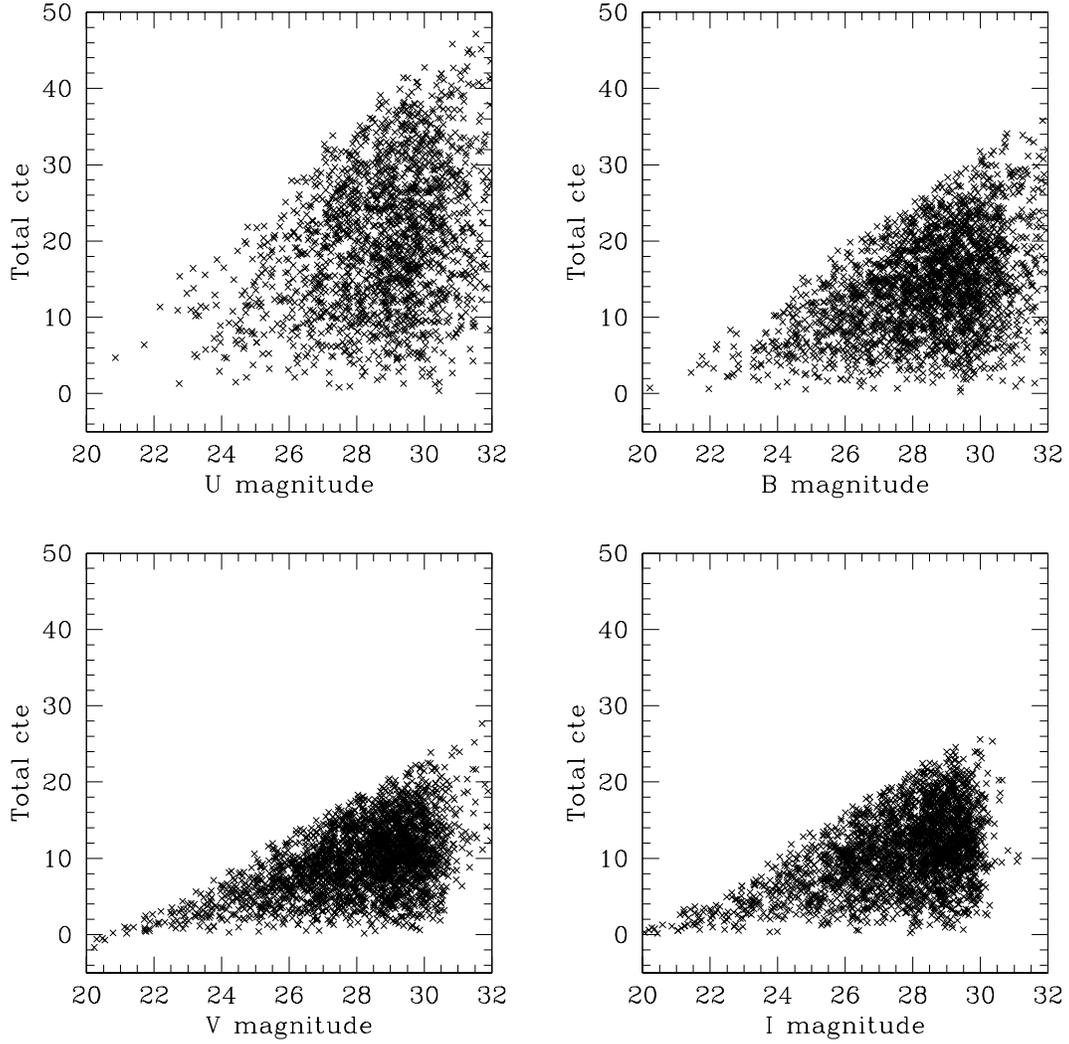,width=6in}}
  \figcaption[Casertano_fig3.ps] {Distribution of the point-source CTE
correction (in percent) suggested by Whitmore {\etal} (1999) for
galaxies in the HDF-S, based on total counts and typical background in
individual images.  The point-source correction most likely
overestimates the CTE-induced photometric error for extended sources,
and is shown here only as indicative of the potential worst-case impact
of the CTE.  No CTE correction has been applied to the HDF-S data. 
\label{fig:whitmore_cte}}
  \end{figure}

 \begin{figure}[p]
 \centerline{\psfig{file=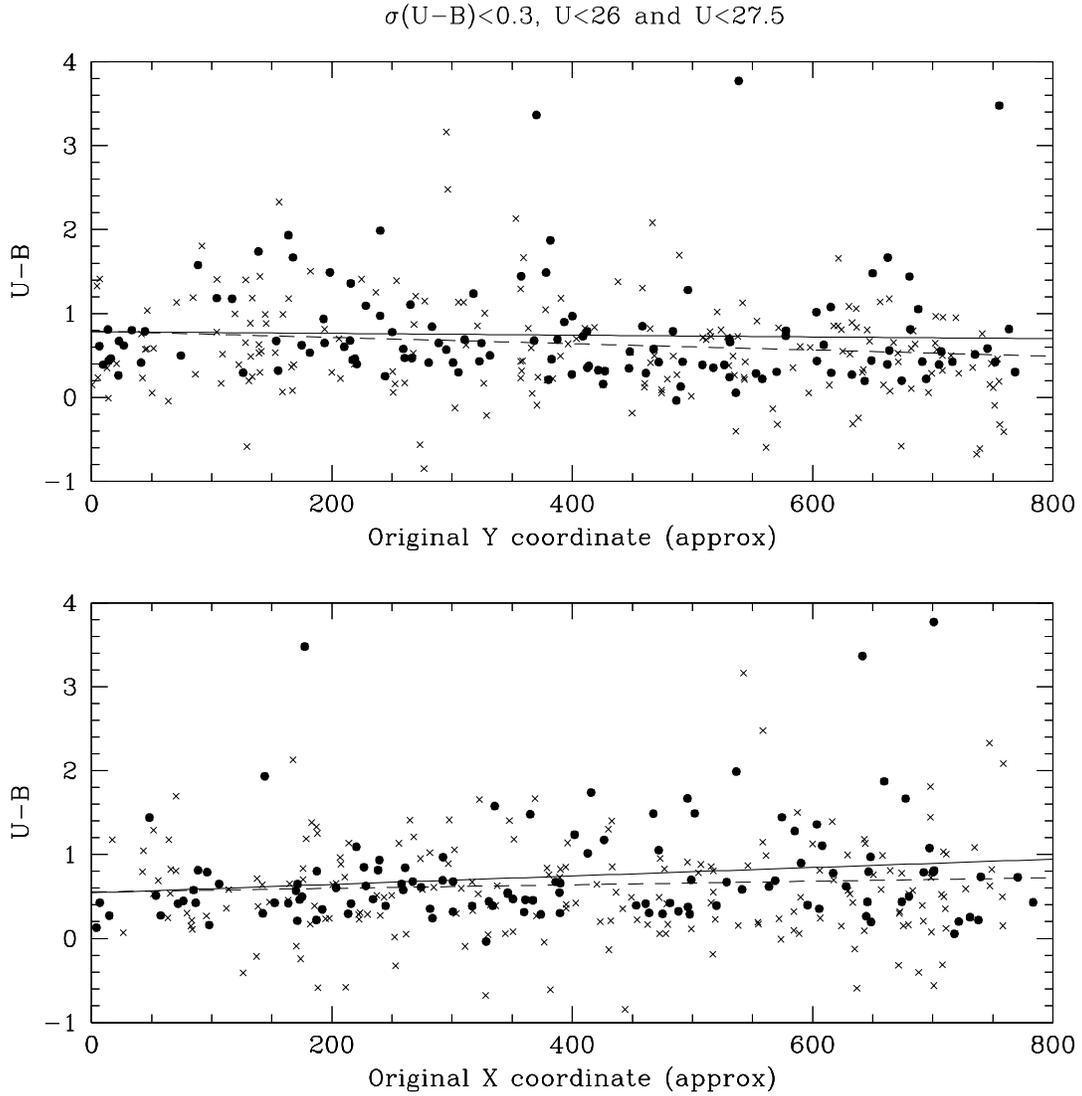,width=6in}}
 \figcaption[Casertano_fig4.ps] {A test for CTE effects: $ U-B $ color
vs.~mean pixel position in the original images, for $ U < 26 $ (solid
dots) and $ 26 < U < 27.5 $ (crosses).  Linear regressions are shown by
solid lines for the dots, and dashed lines for the crosses.  There is no
clear evidence for the position-dependent color variations that could be
ascribed to CTE.\label{fig:umb_cte}}
  \end{figure}

 \begin{figure}[p]
 \centerline{\psfig{file=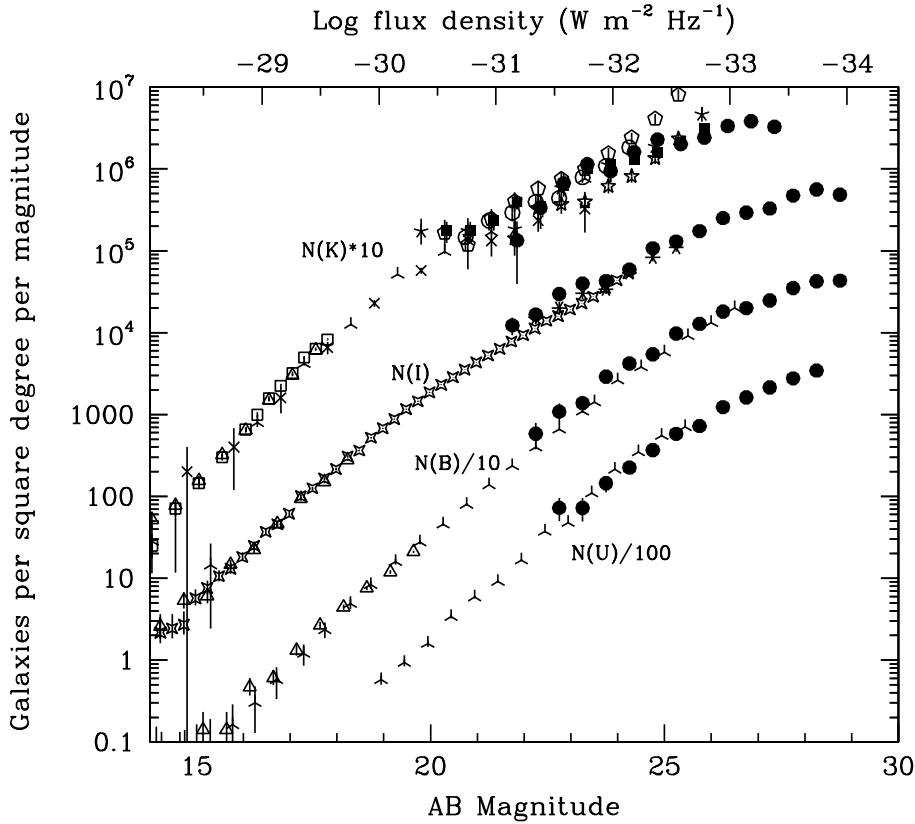,height=5in}}
  \figcaption[Casertano_fig5.ps] { Galaxy counts from the Hubble deep
fields (HDF) and other surveys.  The HDF galaxy counts ({\it solid
symbols}) use isophotal magnitudes and have not been corrected for
incompleteness.  These corrections will tend to steepen the counts at
the faint end, but in a model-dependent way.  For the $K$-band, a color
correction of $-0.4$ mag has been applied to the NICMOS F160W band
magnitudes.  The HDF-N counts from Thompson et al (1999) ({\it filled
squares}) and the HDF-S counts from Fruchter {\etal} 2000 ({\it filled
circles}) are shown.  For the $U,B$ and $I$ bands, the HDF counts
(filled circles) are the average of HDF-N (Williams {\etal} 1996) and
HDF-S (this paper) with no color corrections.  The groundbased counts
are from Gardner {\etal} (1993; 1996; open triangles), Mcleod and Rieke
(1995, three-pointed symbols), Huang {\etal} (1997, open squares),
Minezaki {\etal} (1998, crosses), Postman {\etal} (1998, four-pointed
stars), Bershady {\etal} (1998, open pentagons), Moustakas {\etal}
(1997, five-pointed symbols), Djorgovski {\etal} (1995, five-pointed
stars), and Lilly, Cowie \& Gardner (1991, six-pointed symbols).
\label{fig:countsonly}}
 \end{figure}

 \begin{figure}[p]
 \centerline{\psfig{file=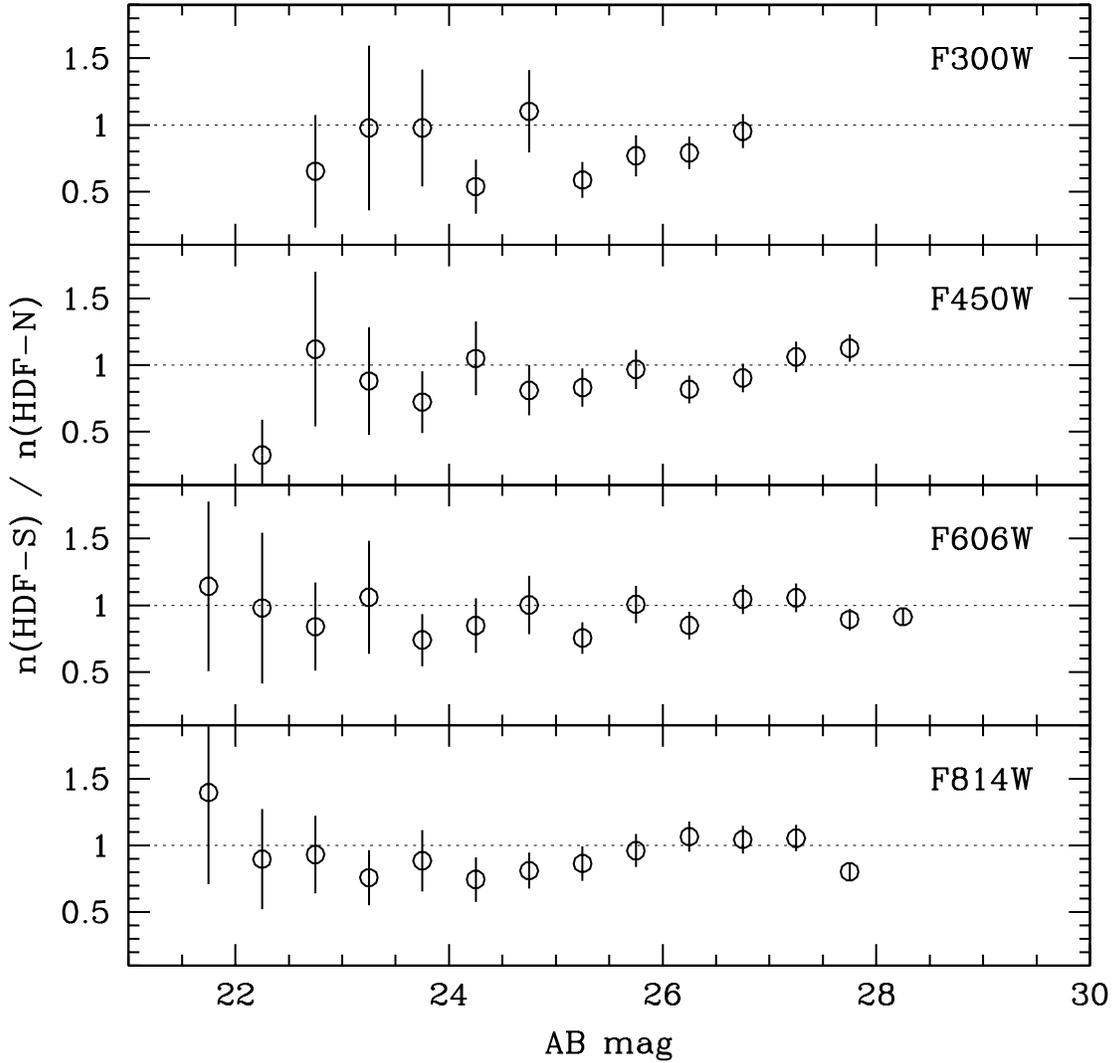,width=6in}}
  \figcaption[Casertano_fig6.ps] {Ratio between number counts in HDF-S
and HDF-N.  The ratios are plotted down to roughly the 10$\sigma$ limit
for galaxy detection.  The slight differences in the depth of the HDF-N
and HDF-S images, and the slight variations in sensitivity across the
field introduce significant corrections at fainter magnitudes, but these
corrections are negligible to the limits shown.\label{fig:ratioall}}
  \end{figure}

 \begin{figure}[p]
 \centerline{\psfig{file=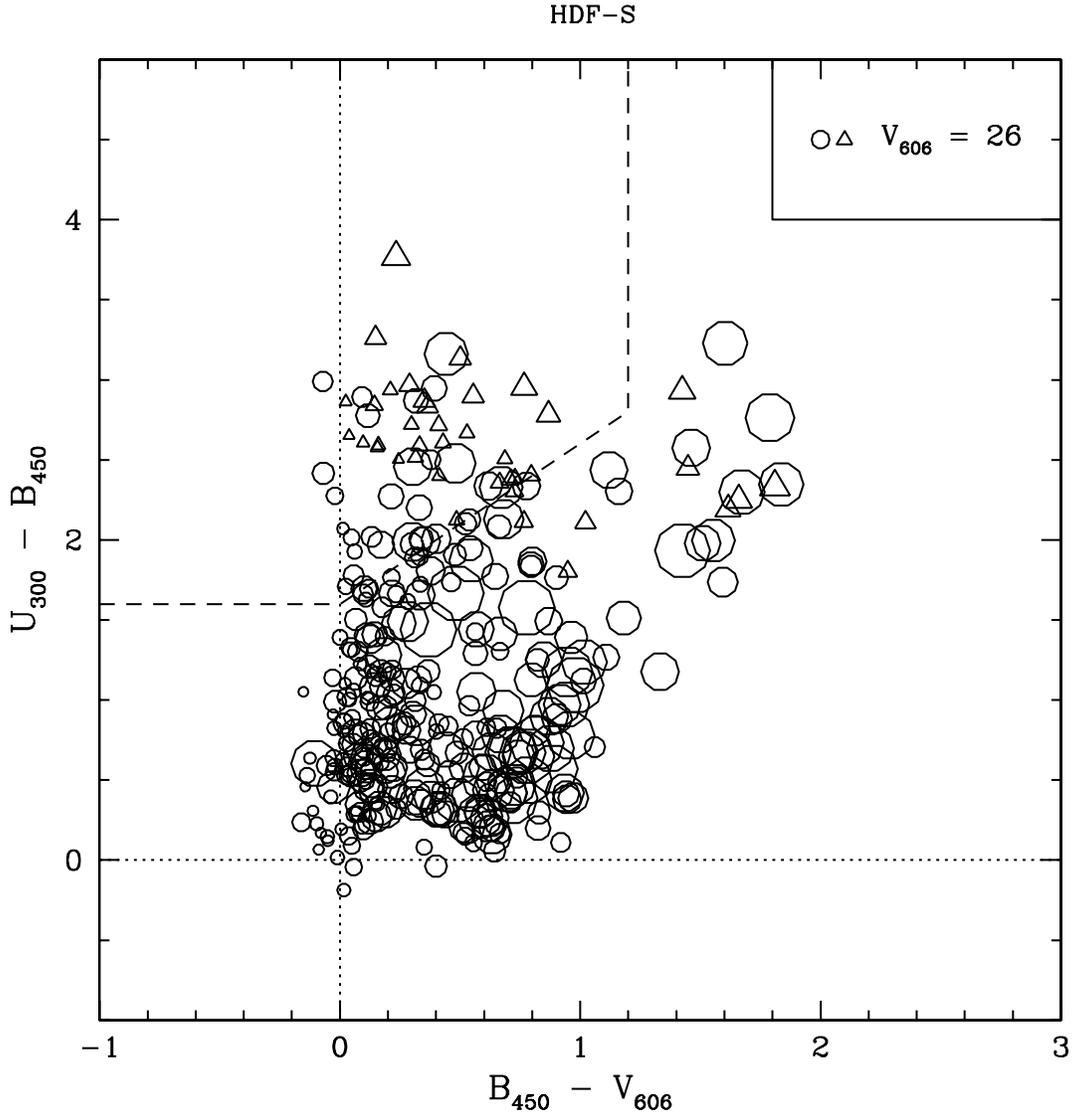,width=6in}}
 \figcaption[Casertano_fig7.ps] {Color-color plot highlighting the
candidate U-band dropouts (above dashed line) in HDF-S.  Circles
indicate measurements; triangles indicate non-detections in $ U_{300} $,
thus lower limits in the ordinate $ U_{300} - B_{450} $.  The size of
the symbols scales with the magnitude in F814W.\label{fig:UdropplotS}}
  \end{figure}

 \begin{figure}[p]
 \centerline{\psfig{file=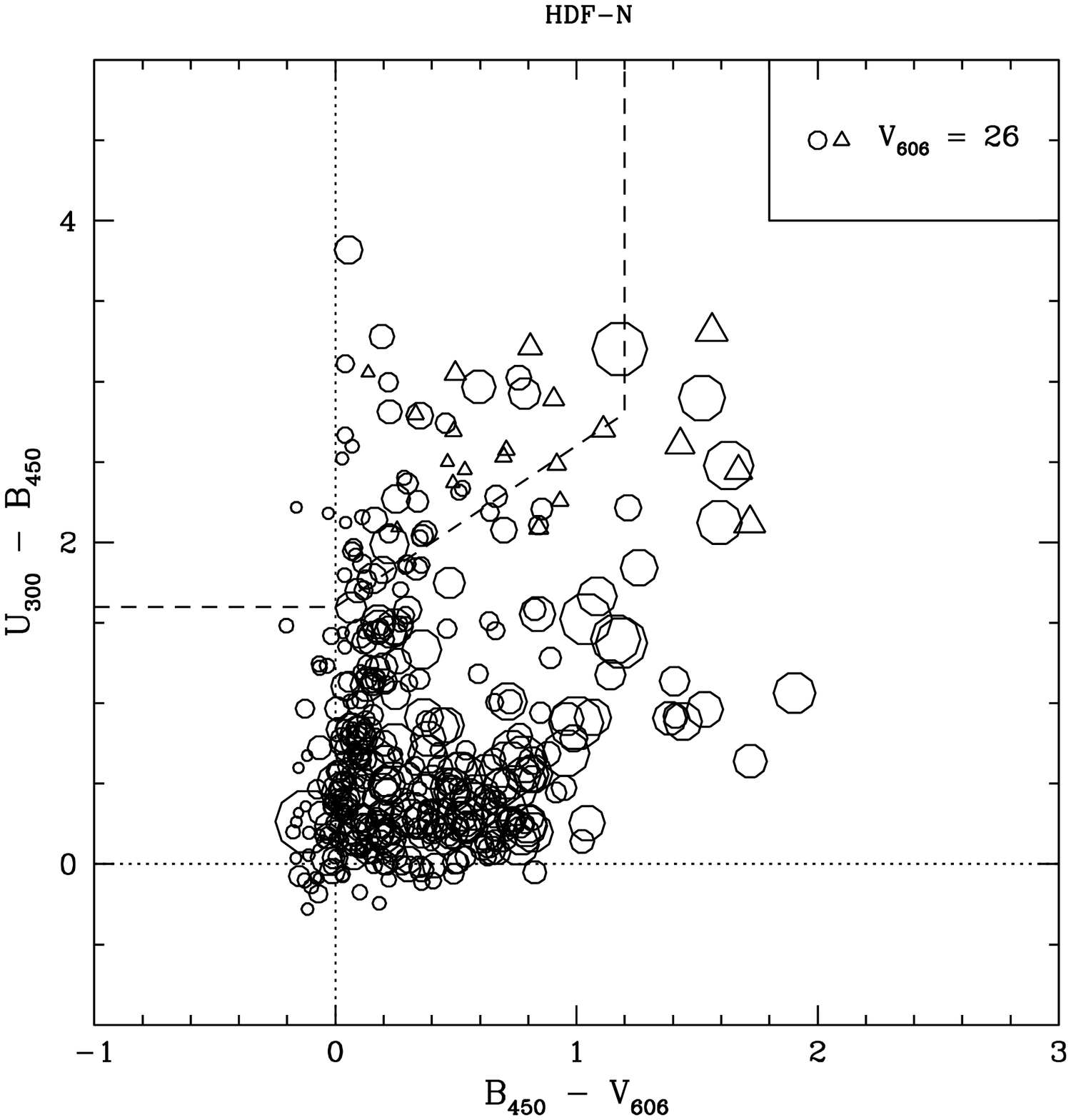,width=6in}}
 \figcaption[Casertano_fig8.ps] {Same as Figure~\ref{fig:UdropplotS},
but for HDF-N.\label{fig:UdropplotN}}
  \end{figure}

 \begin{figure}[p]
 \centerline{\psfig{file=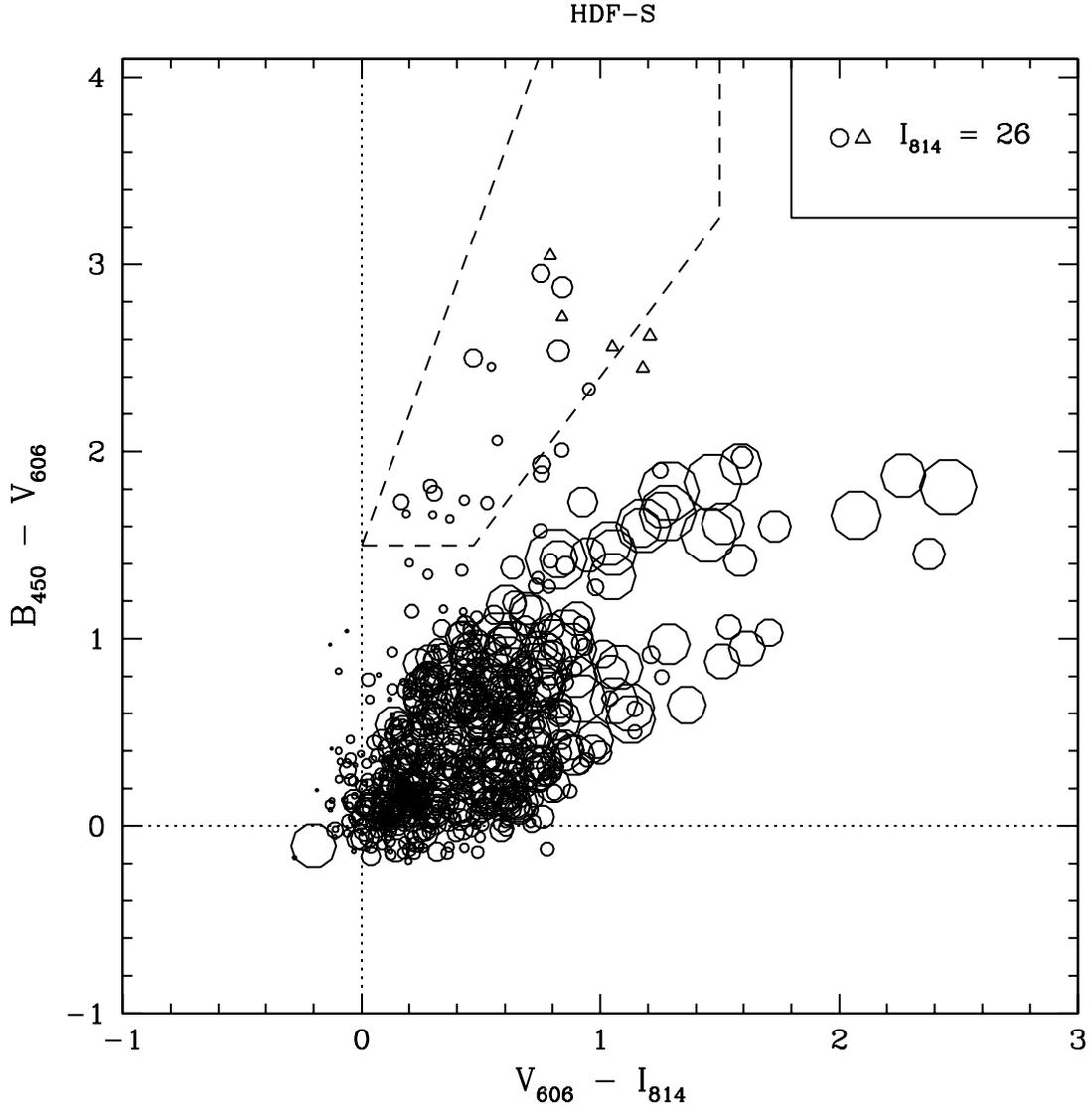,width=6in}}
  \figcaption[Casertano_fig9.ps] {Color-color plot highlighting the
candidate B-band dropouts in HDF-S (within the region bounded by the
dashed line).  Circles indicate measurements; triangles indicate
non-detections in $ B_{450} $, thus lower limits in the ordinate $
B_{450} - V_{606} $.  The size of the symbols scales with the magnitude
in F814W.  \label{fig:BdropplotS}}
  \end{figure}

 \begin{figure}[p]
 \centerline{\psfig{file=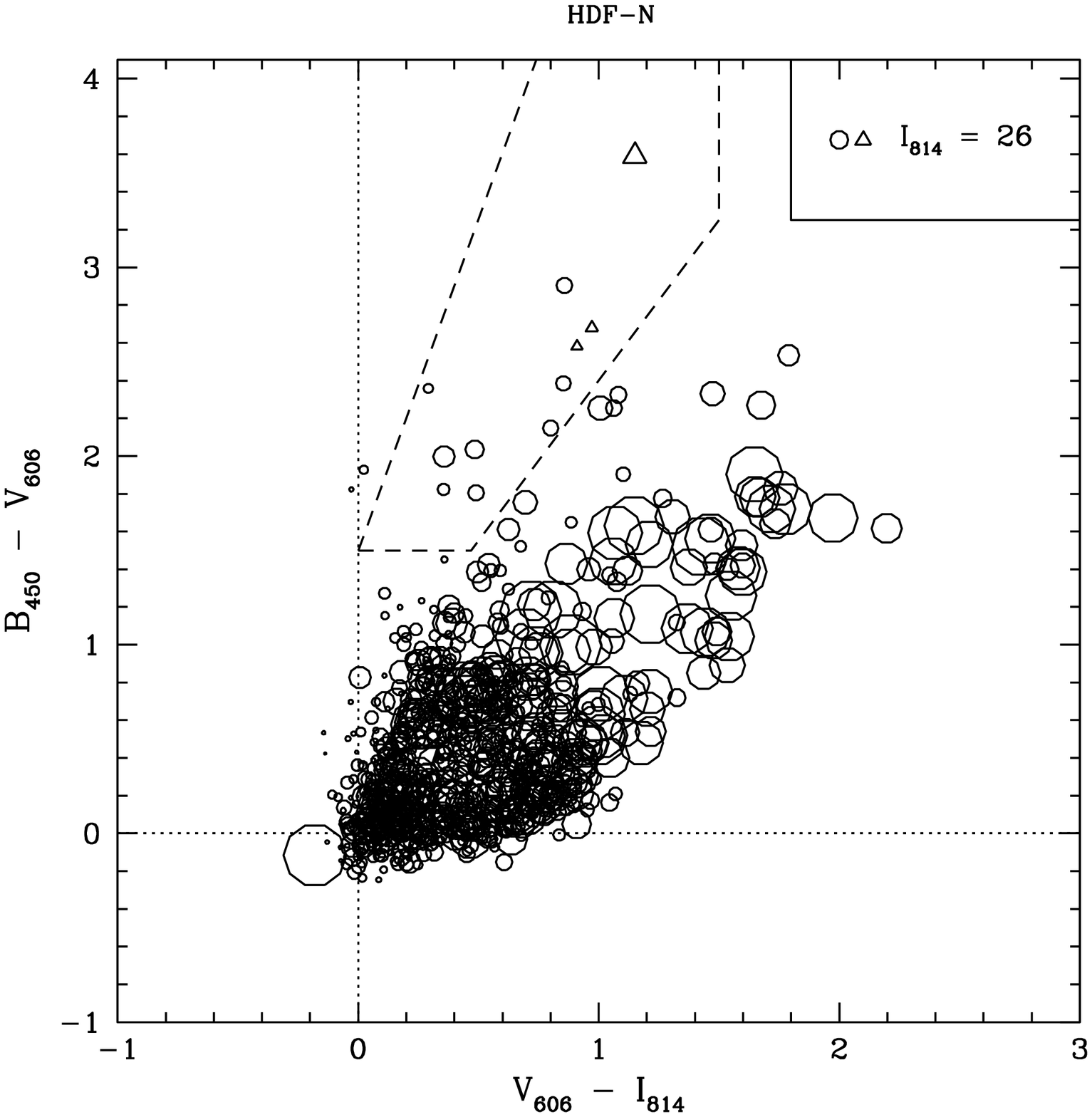,width=6in}}
 \figcaption[Casertano_fig10.ps] {Same as Figure~\ref{fig:BdropplotS},
but for HDF-N.\label{fig:BdropplotN}}
  \end{figure}

 \begin{figure}[p]
 \centerline{\psfig{file=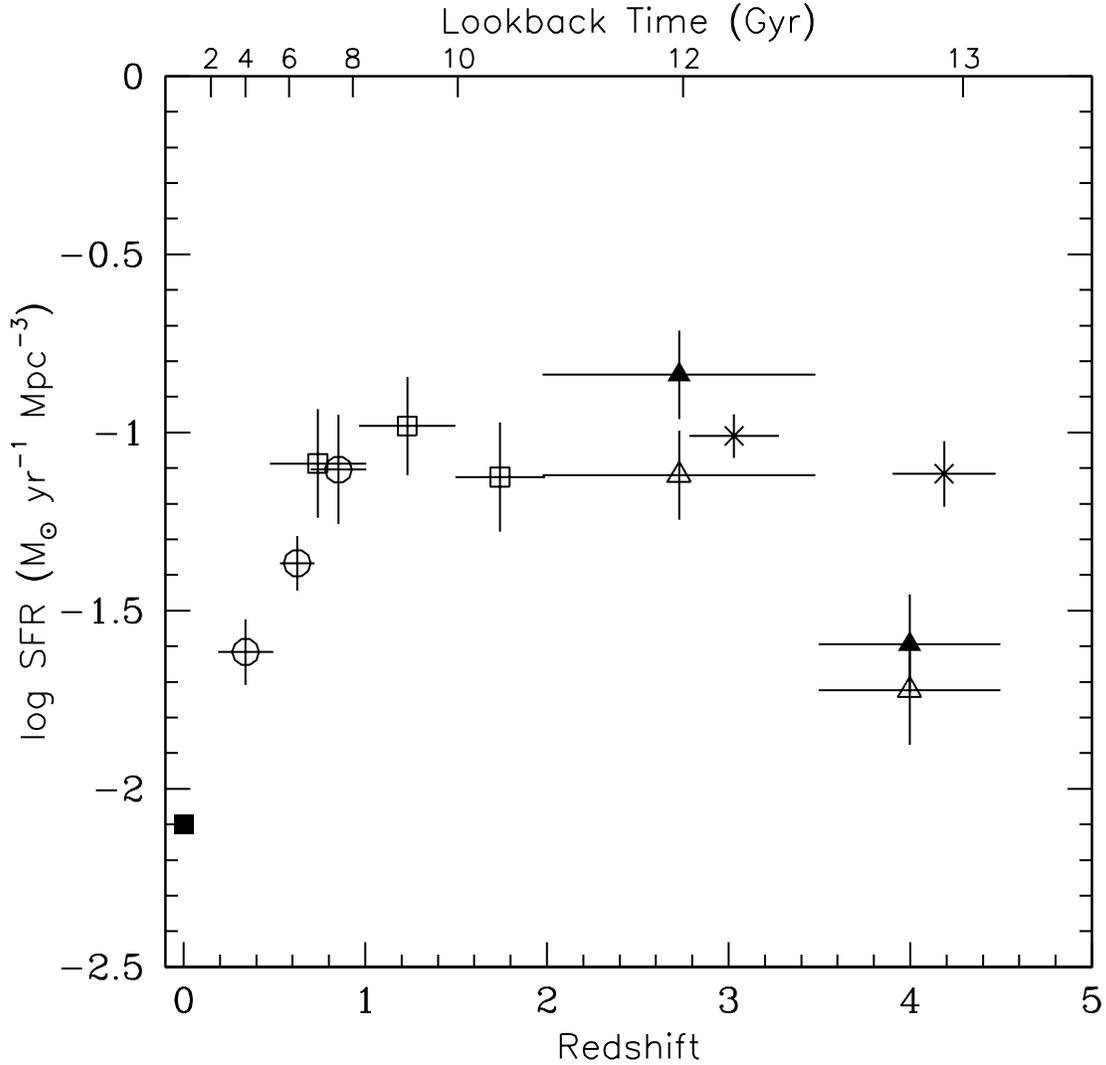,width=6in}}
  \figcaption[Casertano_fig11.ps] {Star-formation rate density vs. 
redshift derived from ultraviolet luminosity density.  The $z > 2$
points are from Lyman-break objects in the HDF-N (open triangles),
in the HDF-S (filled triangles), and in the Steidel {\etal} (1999)
ground-based survey ($\times$ symbols).  The luminosity density
has been determined by integrating over the luminosity function and
correcting for extinction following the prescription of Steidel {\etal}
(1999).  Distances and volumes are computed using the cosmological
parameters $h,\Omega_{\rm m}, \Omega_\Lambda, \Omega_{\rm tot} = 0.65,
0.3, 0.7, 1.0.$ Possible contributions from far-IR and sub-millimeter
sources are not included.  Also not included are the upward revisions of
the $z < 1$ star-formation densities suggested by Tresse \& Maddox
(1998) and Cowie {\etal} (1999).  For the lower-redshift points, the
open squares are from HDF photometric redshifts by Connolly et al
(1997), the open circles are from Lilly {\etal} (1996), the
solid square is from the H$\alpha$ survey of Gallego {\etal}
(1995), and the solid circle from Sullivan {\etal} (2000).
\label{fig:sfr_z1}}
  \end{figure}

\clearpage

 \begin{figure}[p]
 \centerline{\psfig{file=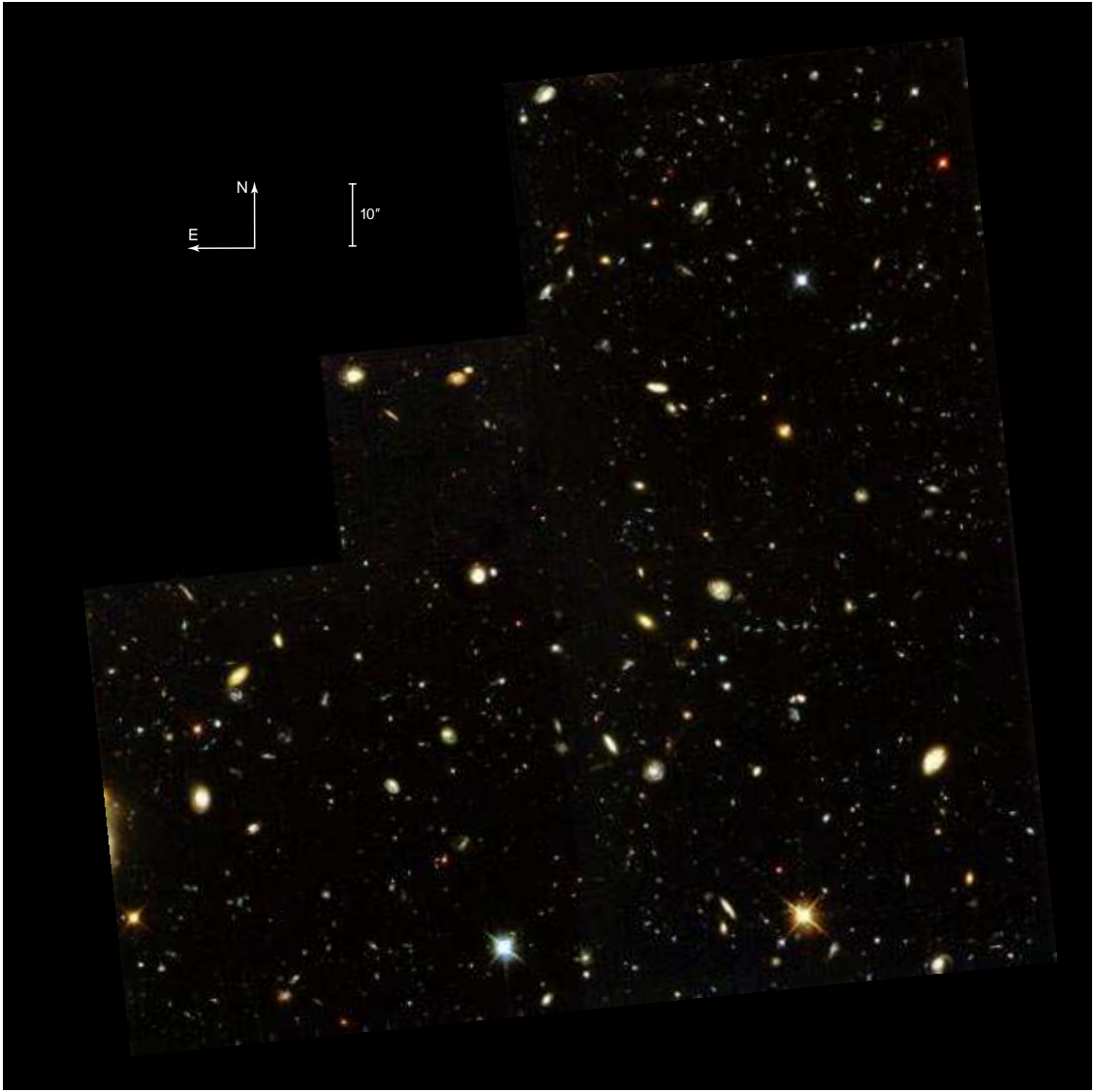,height=6in,bbllx=10pt,bblly=100pt,bburx=600pt,bbury=690pt}}
 \figcaption[Casertano_fig12.ps] { Three-color composite image of the WFPC2
field, using F450W as blue, F606W as green, and F814W as red.}
 \end{figure}
 
 \begin{figure}[p]
 \centerline{\psfig{file=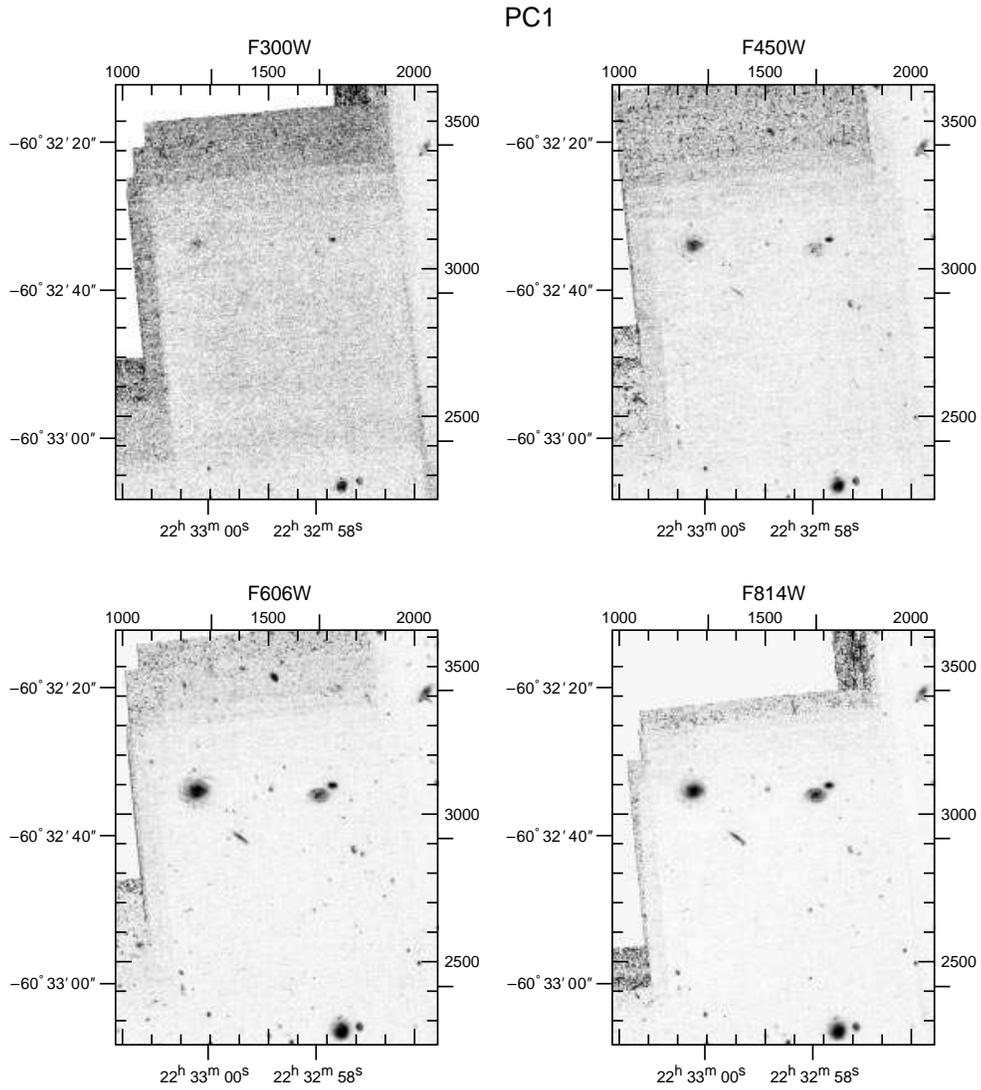,height=6in,bbllx=0pt,bblly=40pt,bburx=600pt,bbury=730pt}}
 \figcaption[Casertano_fig13.ps] { The NE quadrant of the WFPC2 field,
roughly corresponding to the area covered by the Planetary Camera, in
each of the four filters.  The labels on the top and right of each panel
(inner tickmarks) refer to pixel coordinates in the combined image; the
labels on the bottom and left (outer tickmarks) to J2000 celestial
coordinates.}
 \end{figure}
 
 \begin{figure}[p]
 \centerline{\psfig{file=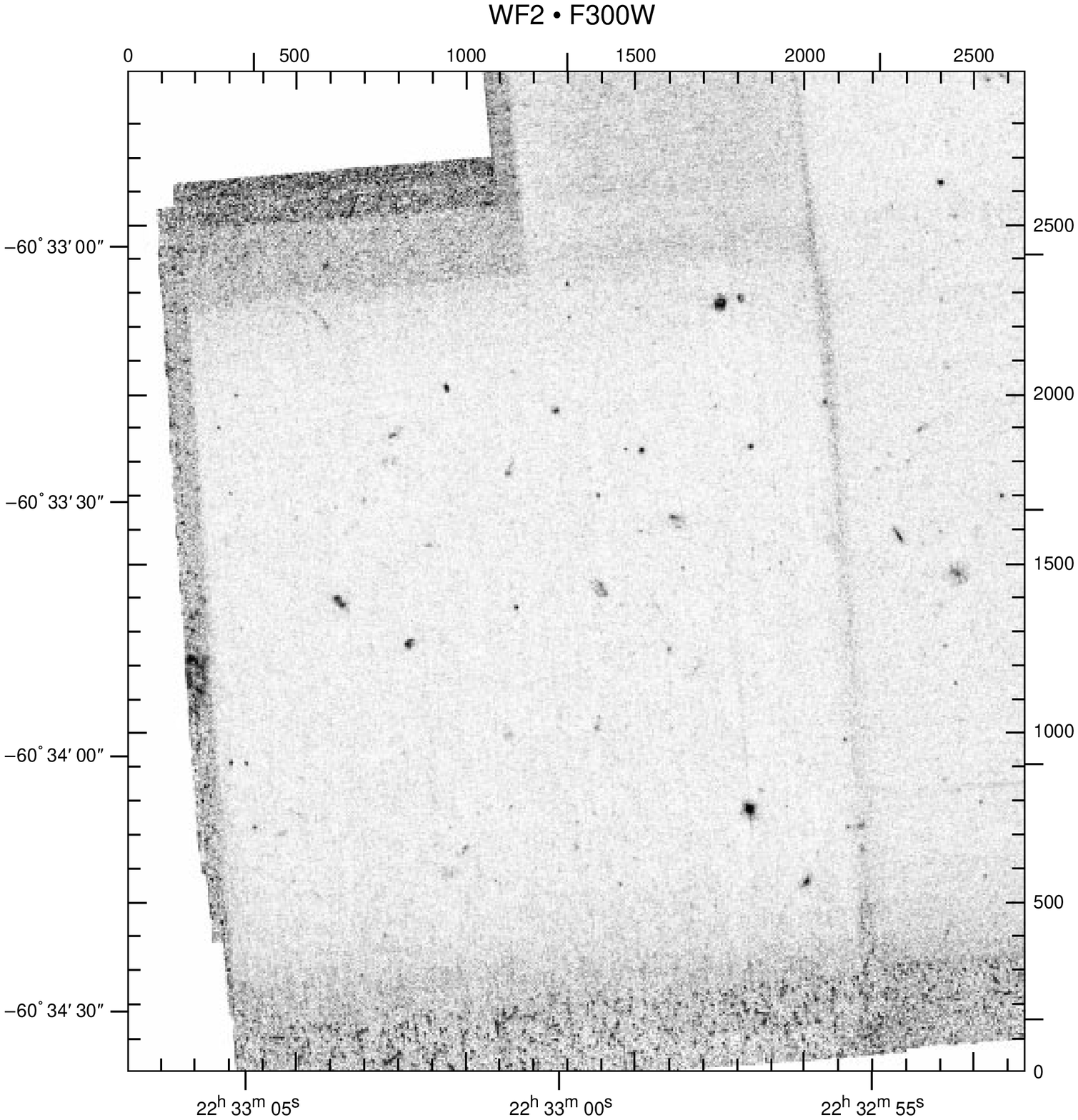,height=6in,bbllx=50pt,bblly=70pt,bburx=560pt,bbury=680pt}}
 \figcaption[Casertano_fig14.ps] { The SE quadrant of the WFPC2 field,
roughly corresponding to the area covered by the Wide Field Camera 2, in
F300W.}
 \end{figure}
 
 \begin{figure}[p]
 \centerline{\psfig{file=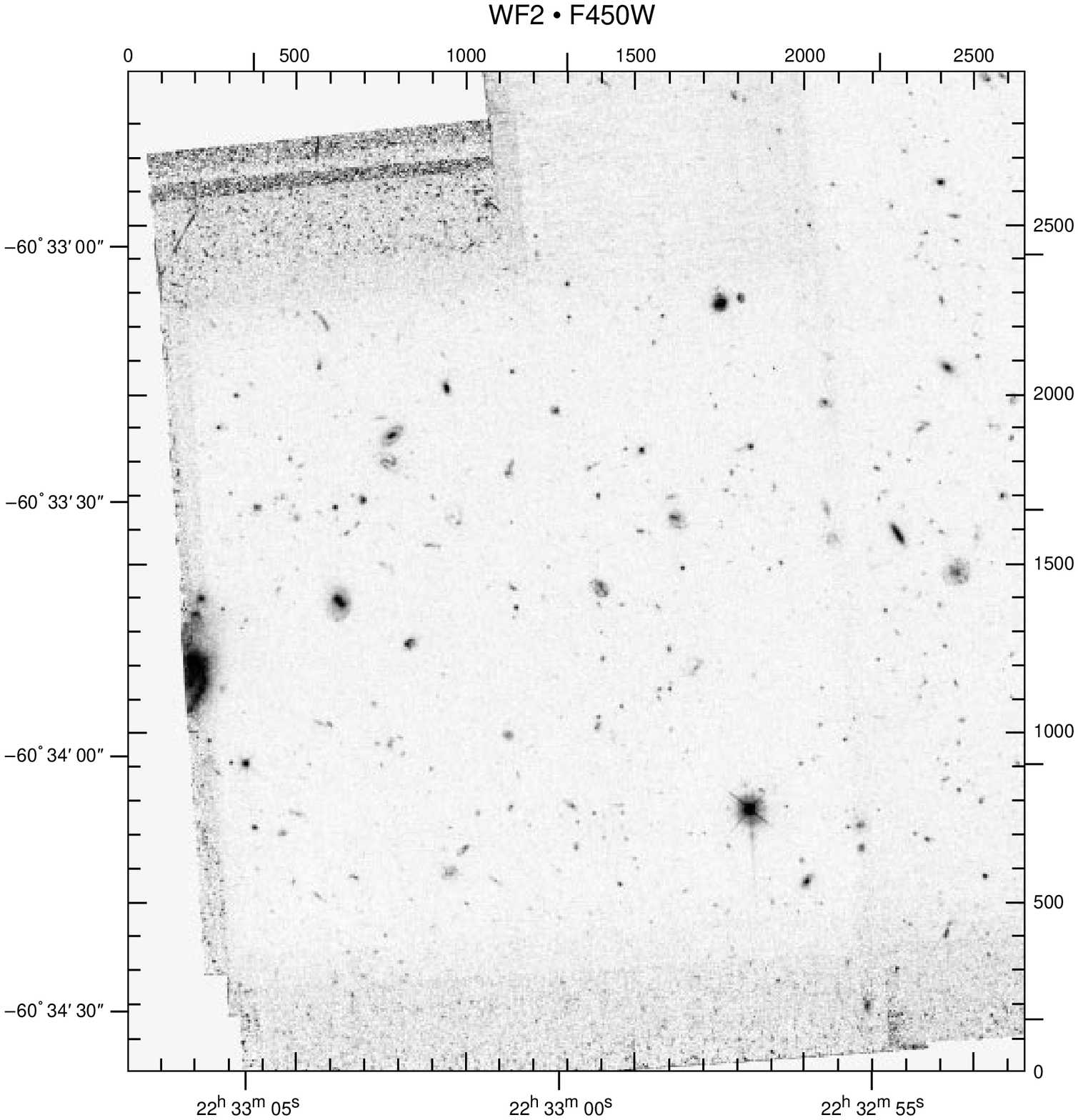,height=6in,bbllx=50pt,bblly=70pt,bburx=560pt,bbury=680pt}}
 \figcaption[Casertano_fig15.ps] { The SE quadrant of the WFPC2 field,
roughly corresponding to the area covered by the Wide Field Camera 2, in
F450W.}
 \end{figure}
 
 \begin{figure}[p]
 \centerline{\psfig{file=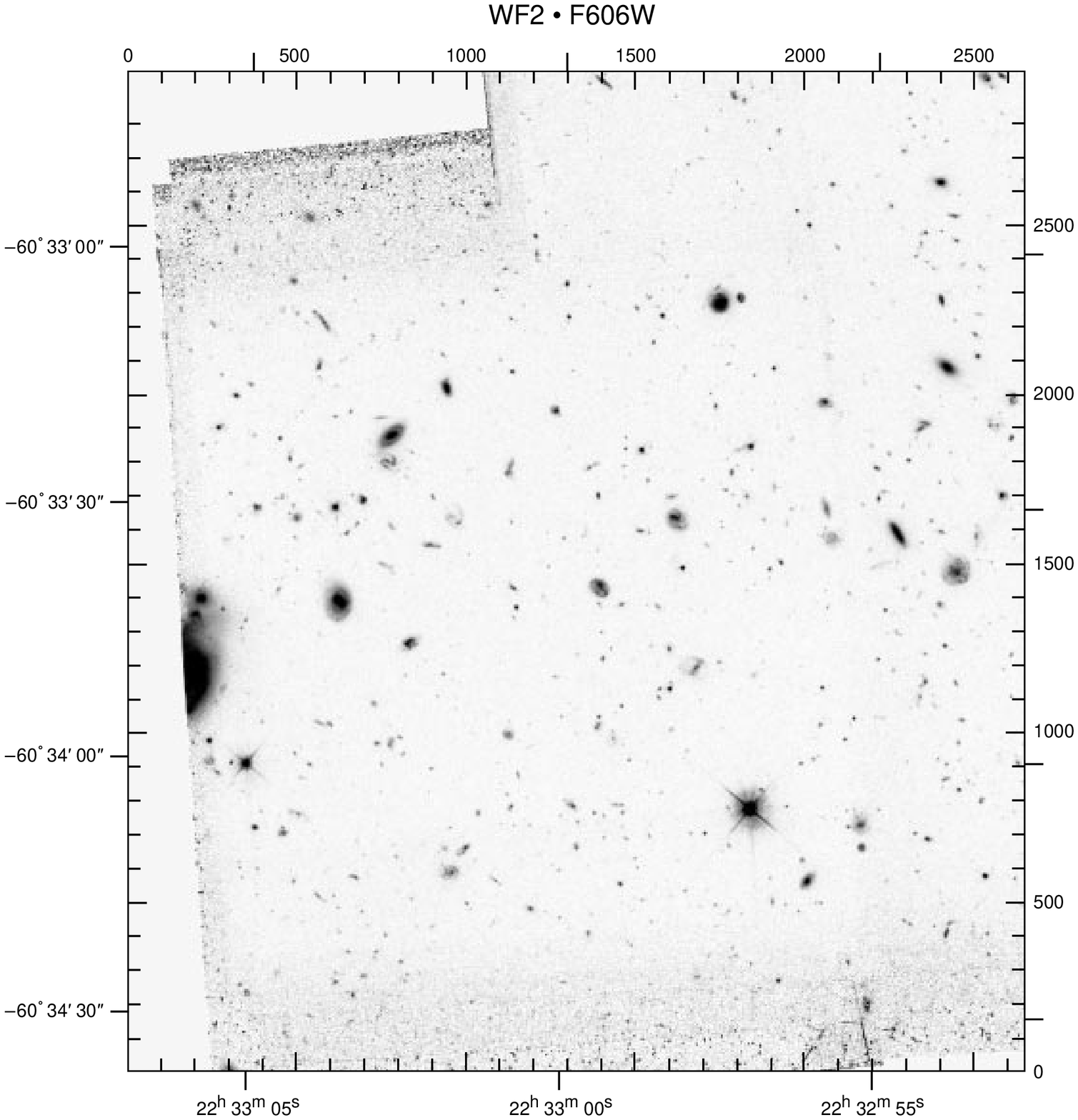,height=6in,bbllx=50pt,bblly=70pt,bburx=560pt,bbury=680pt}}

 \figcaption[Casertano_fig16.ps] { The SE quadrant of the WFPC2 field,
roughly corresponding to the area covered by the Wide Field Camera 2, in
F606W.}
 \end{figure}
 
 \begin{figure}[p]
 \centerline{\psfig{file=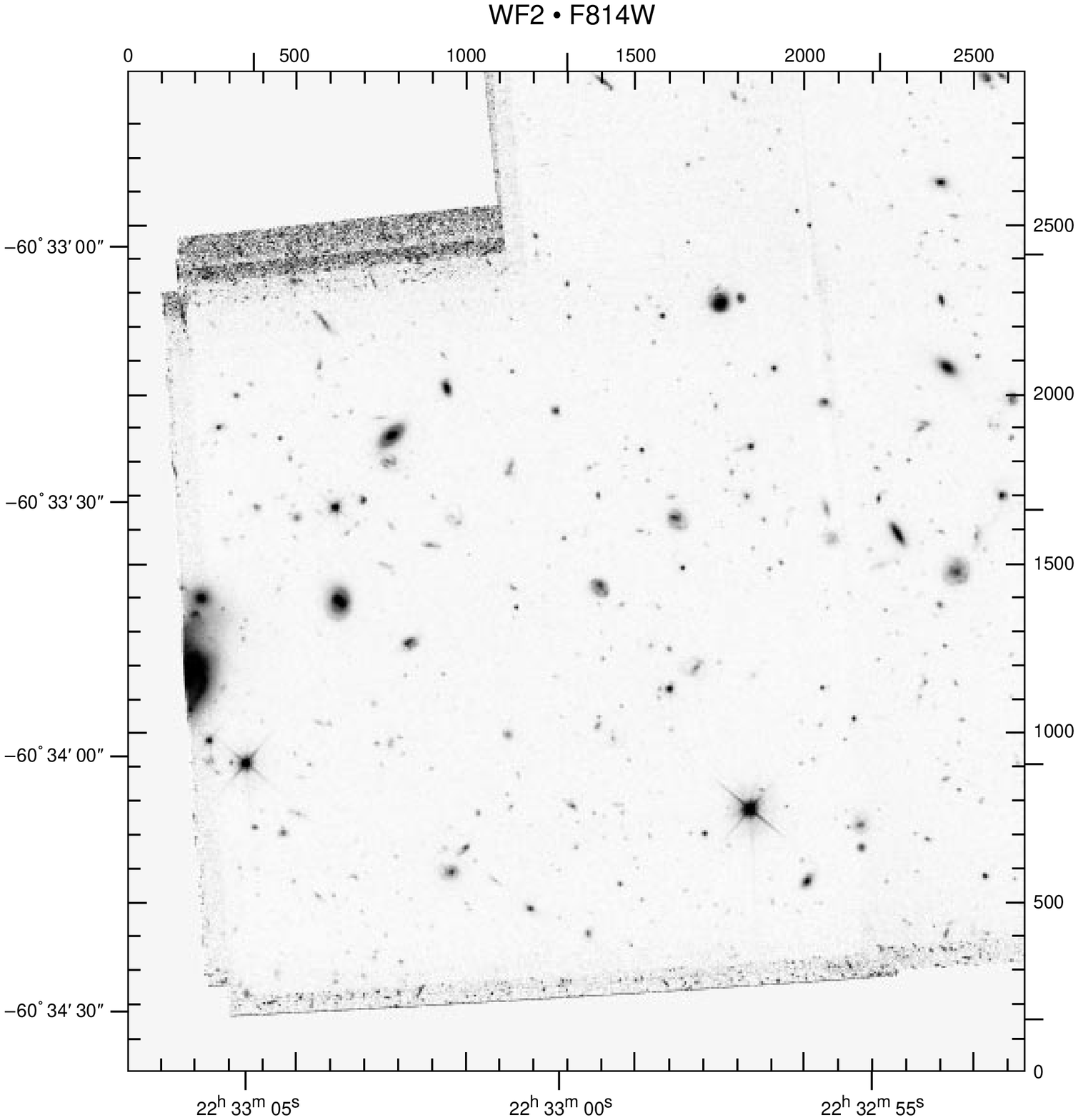,height=6in,bbllx=50pt,bblly=70pt,bburx=560pt,bbury=680pt}}
 \figcaption[Casertano_fig17.ps] { The SE quadrant of the WFPC2 field,
roughly corresponding to the area covered by the Wide Field Camera 2, in
F814W.}
 \end{figure}
 
 \begin{figure}[p]
 \centerline{\psfig{file=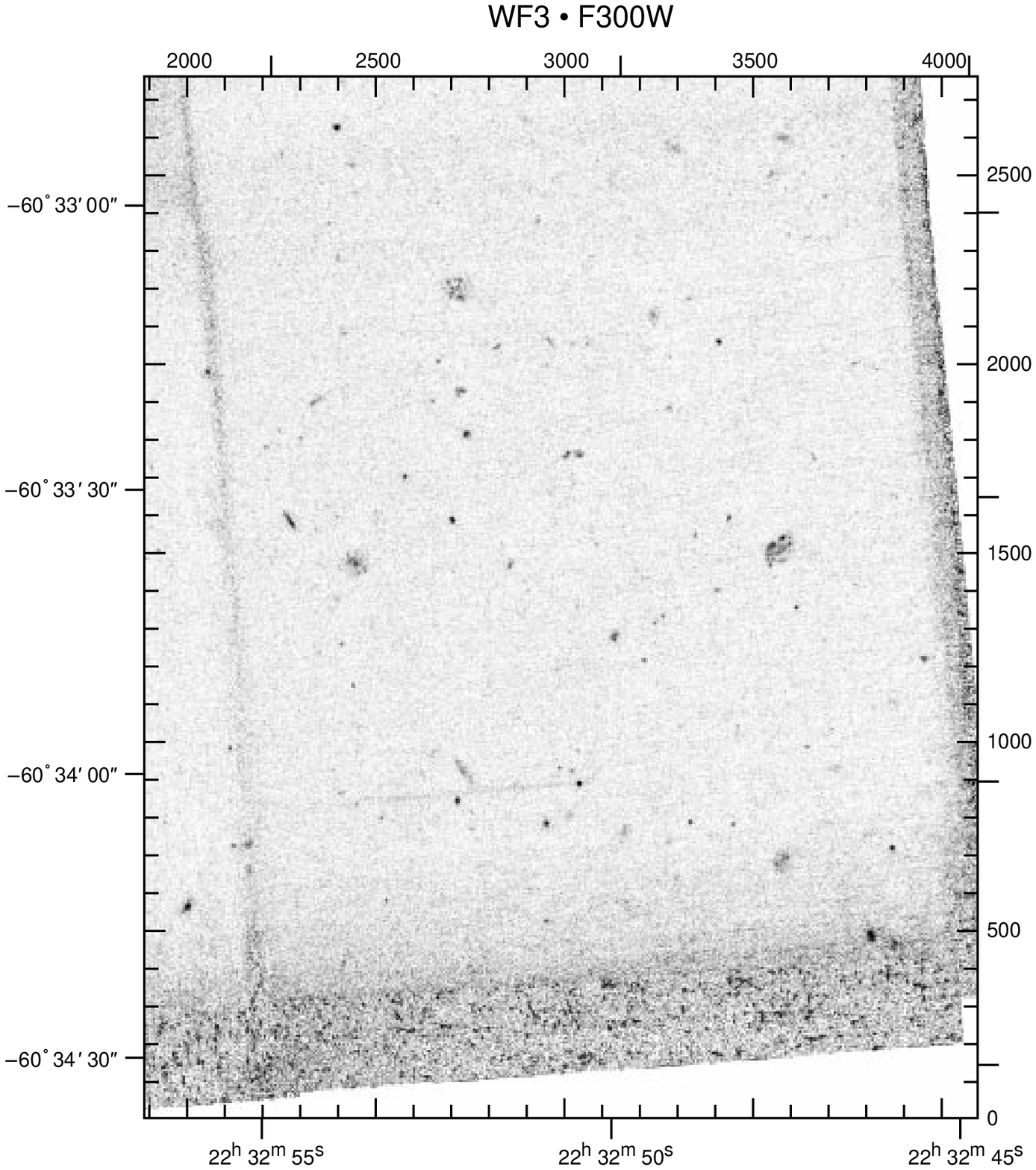,height=6in,bbllx=50pt,bblly=70pt,bburx=560pt,bbury=680pt}}
 \figcaption[Casertano_fig18.ps] { The SW quadrant of the WFPC2 field,
roughly corresponding to the area covered by the Wide Field Camera 3, in
F300W.}
 \end{figure}
 
 \begin{figure}[p]
 \centerline{\psfig{file=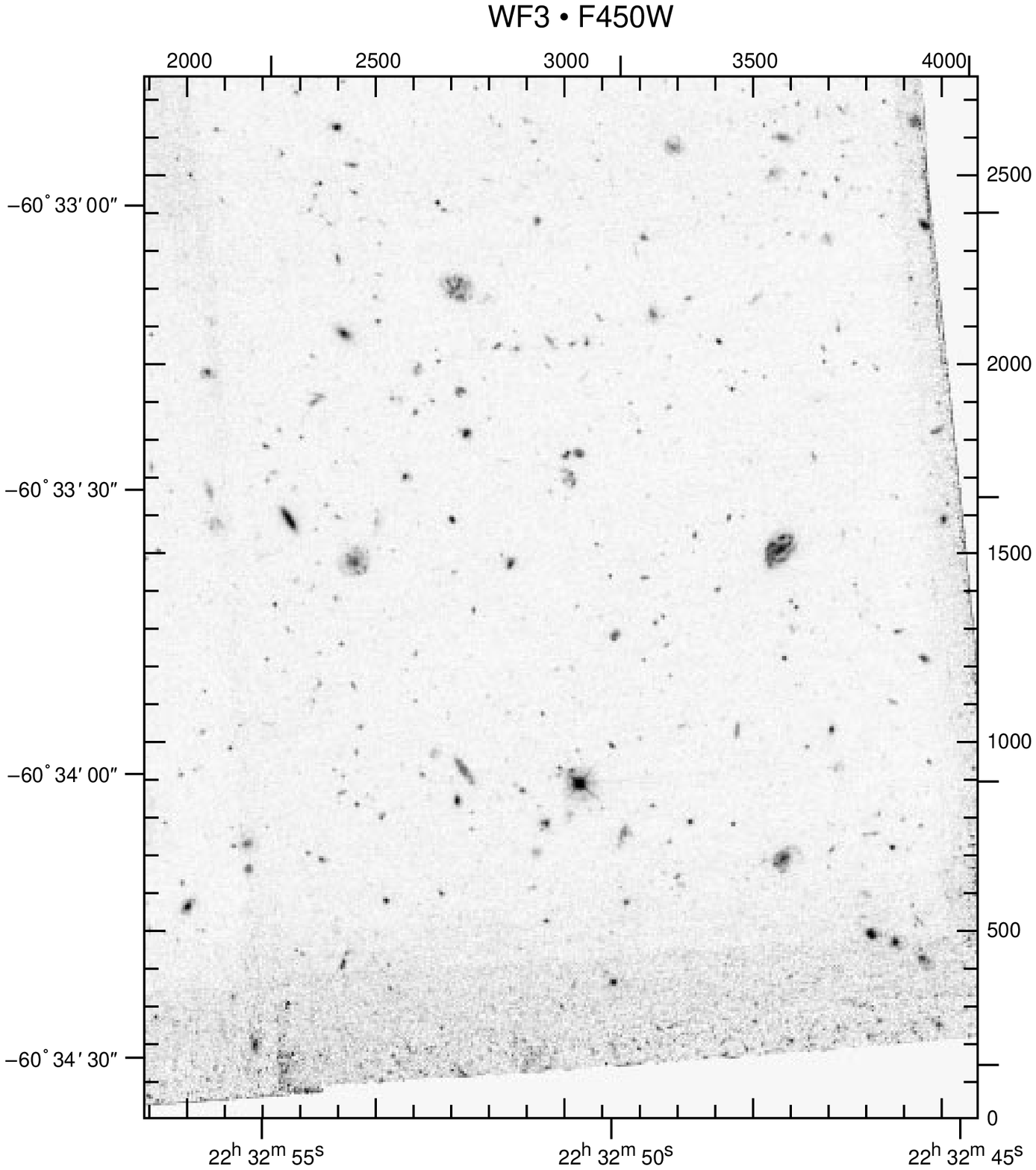,height=6in,bbllx=50pt,bblly=70pt,bburx=560pt,bbury=680pt}}
 \figcaption[Casertano_fig19.ps] { The SW quadrant of the WFPC2 field,
roughly corresponding to the area covered by the Wide Field Camera 3, in
F450W.}
 \end{figure}
 
 \begin{figure}[p]
 \centerline{\psfig{file=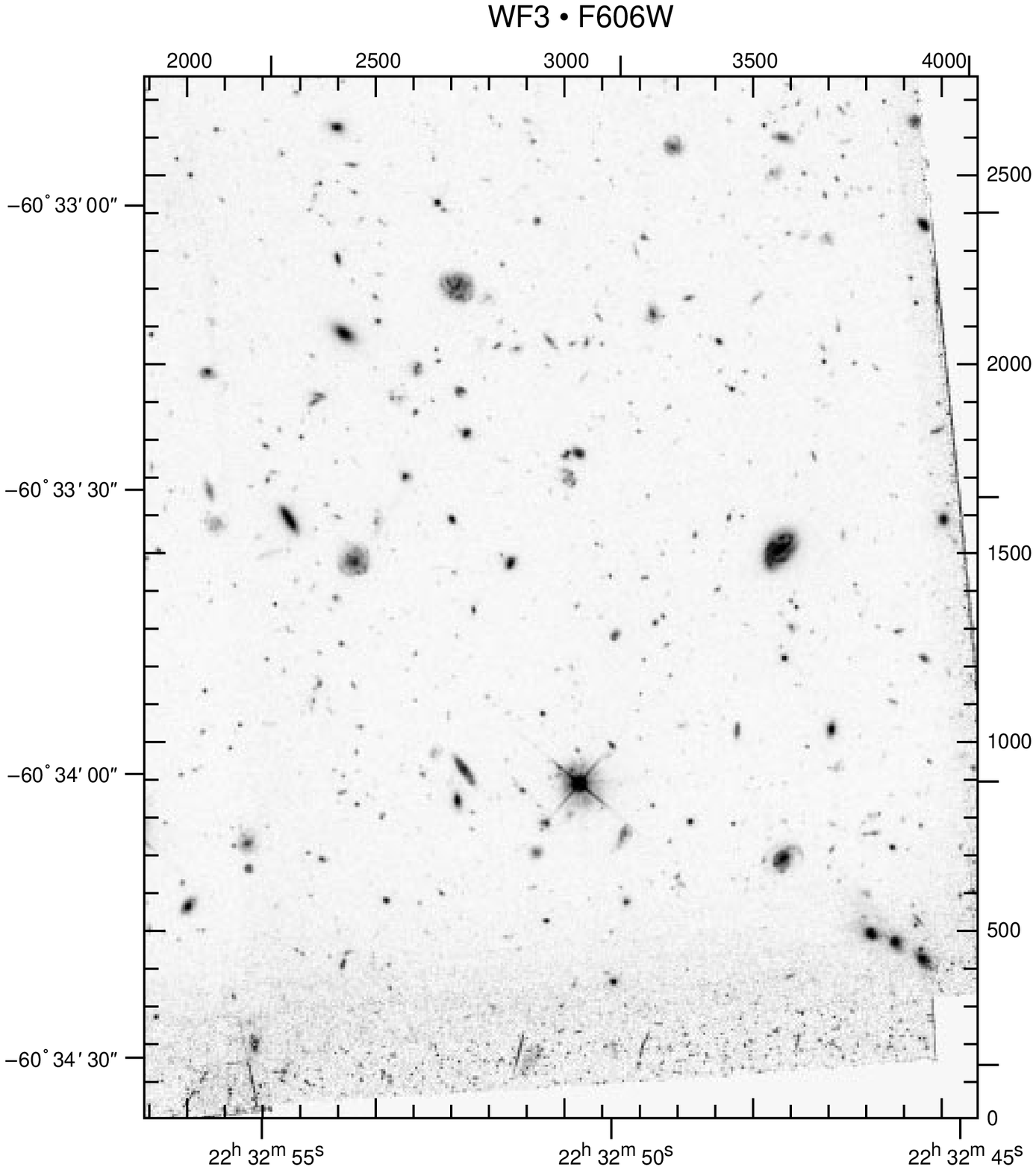,height=6in,bbllx=50pt,bblly=70pt,bburx=560pt,bbury=680pt}}
 \figcaption[Casertano_fig20.ps] { The SW quadrant of the WFPC2 field,
roughly corresponding to the area covered by the Wide Field Camera 3, in
F606W.}
 \end{figure}
 
 \begin{figure}[p]
 \centerline{\psfig{file=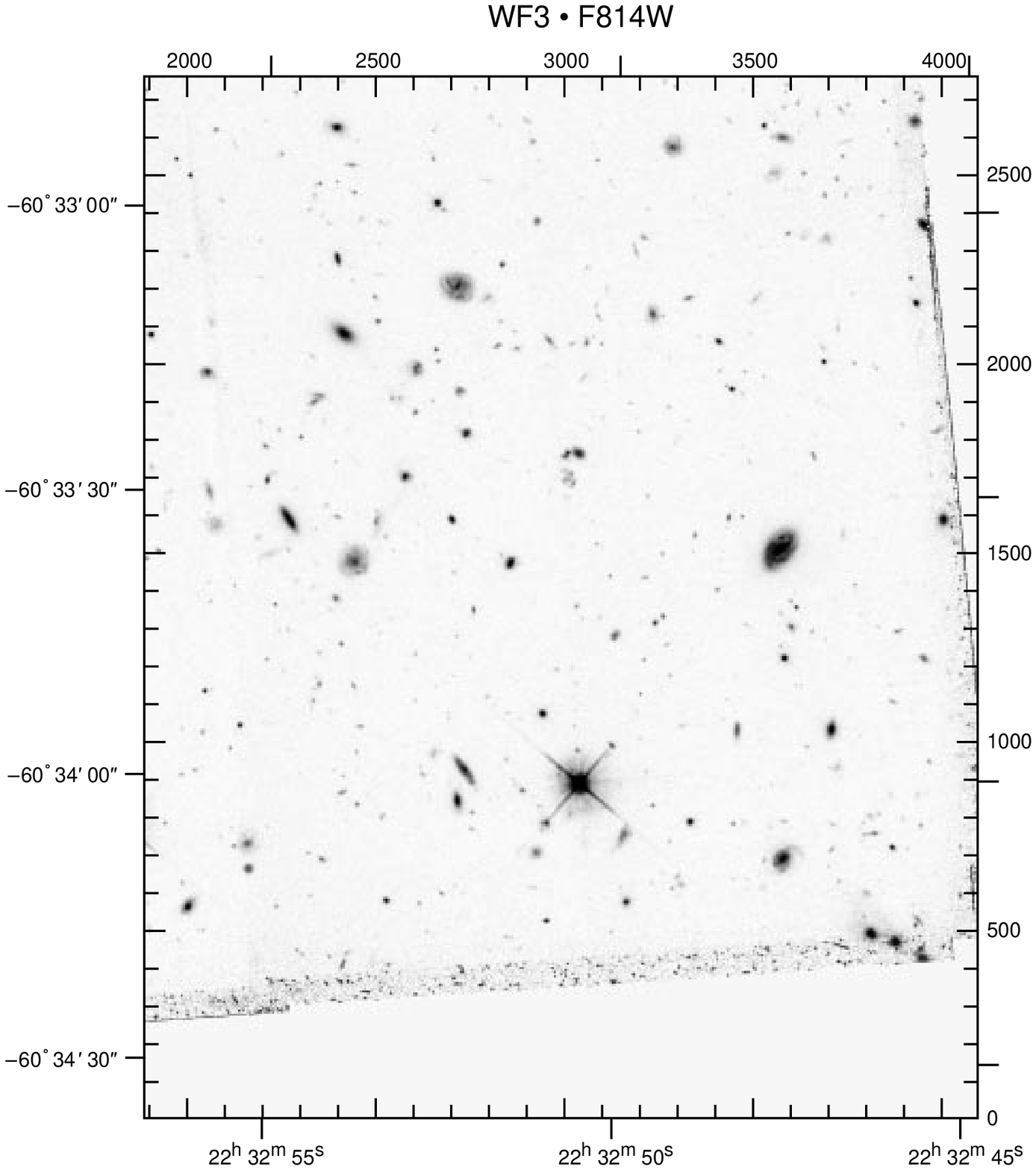,height=6in,bbllx=50pt,bblly=70pt,bburx=560pt,bbury=680pt}}
 \figcaption[Casertano_fig21.ps] { The SW quadrant of the WFPC2 field,
roughly corresponding to the area covered by the Wide Field Camera 3, in
F814W.}
 \end{figure}
 
 \begin{figure}[p]
 \centerline{\psfig{file=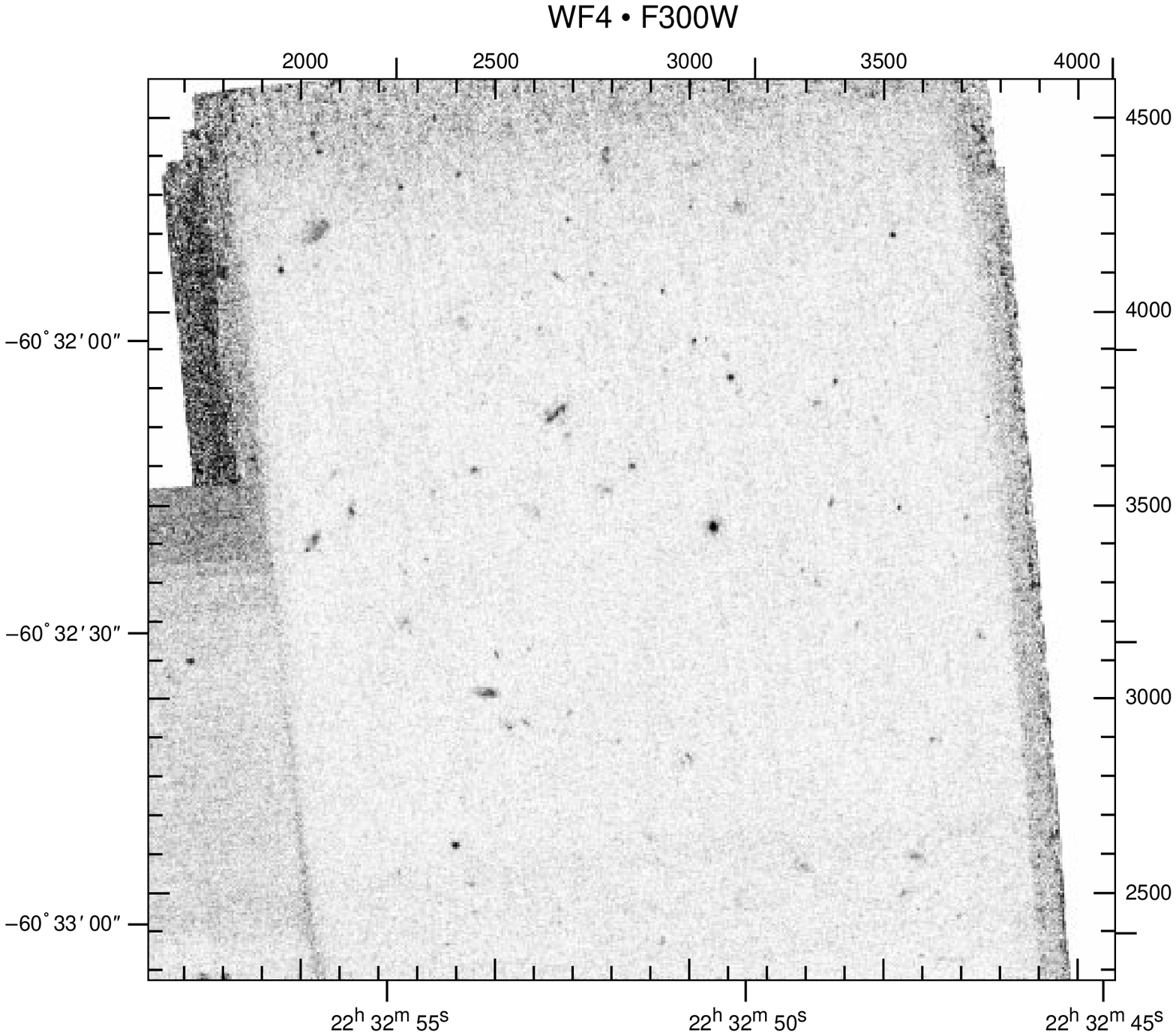,height=6in,bbllx=50pt,bblly=70pt,bburx=560pt,bbury=680pt}}
 \figcaption[Casertano_fig22.ps] { The NW quadrant of the WFPC2 field,
roughly corresponding to the area covered by the Wide Field Camera 4, in
F300W.}
 \end{figure}
 
 \begin{figure}[p]
 \centerline{\psfig{file=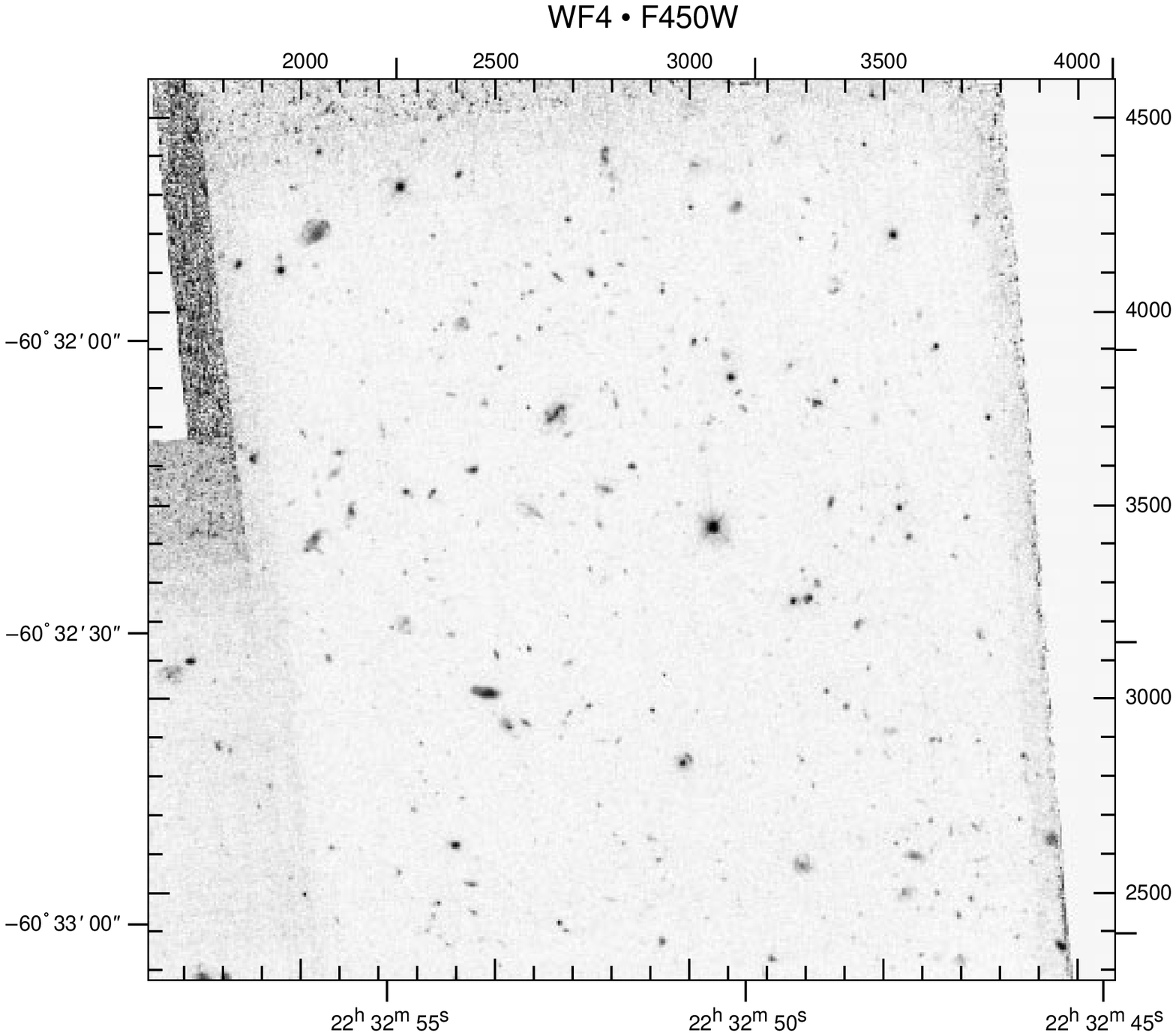,height=6in,bbllx=50pt,bblly=70pt,bburx=560pt,bbury=680pt}}
 \figcaption[Casertano_fig23.ps] { The NW quadrant of the WFPC2 field,
roughly corresponding to the area covered by the Wide Field Camera 4, in
F450W.}
 \end{figure}
 
 \begin{figure}[p]
 \centerline{\psfig{file=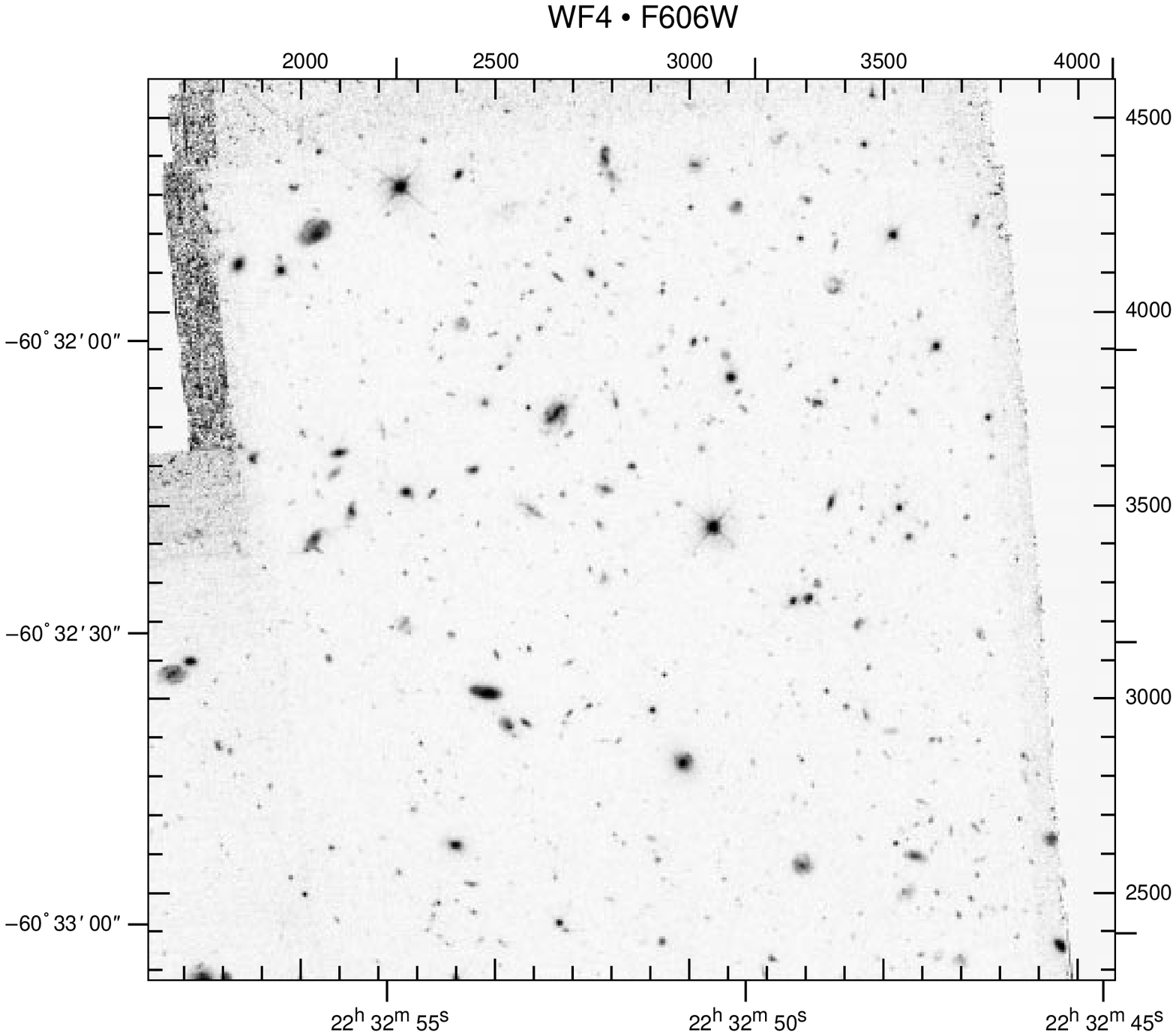,height=6in,bbllx=50pt,bblly=70pt,bburx=560pt,bbury=680pt}}
 \figcaption[Casertano_fig24.ps] { The NW quadrant of the WFPC2 field,
roughly corresponding to the area covered by the Wide Field Camera 4, in
F606W.}
 \end{figure}
 
 \begin{figure}[p]
 \centerline{\psfig{file=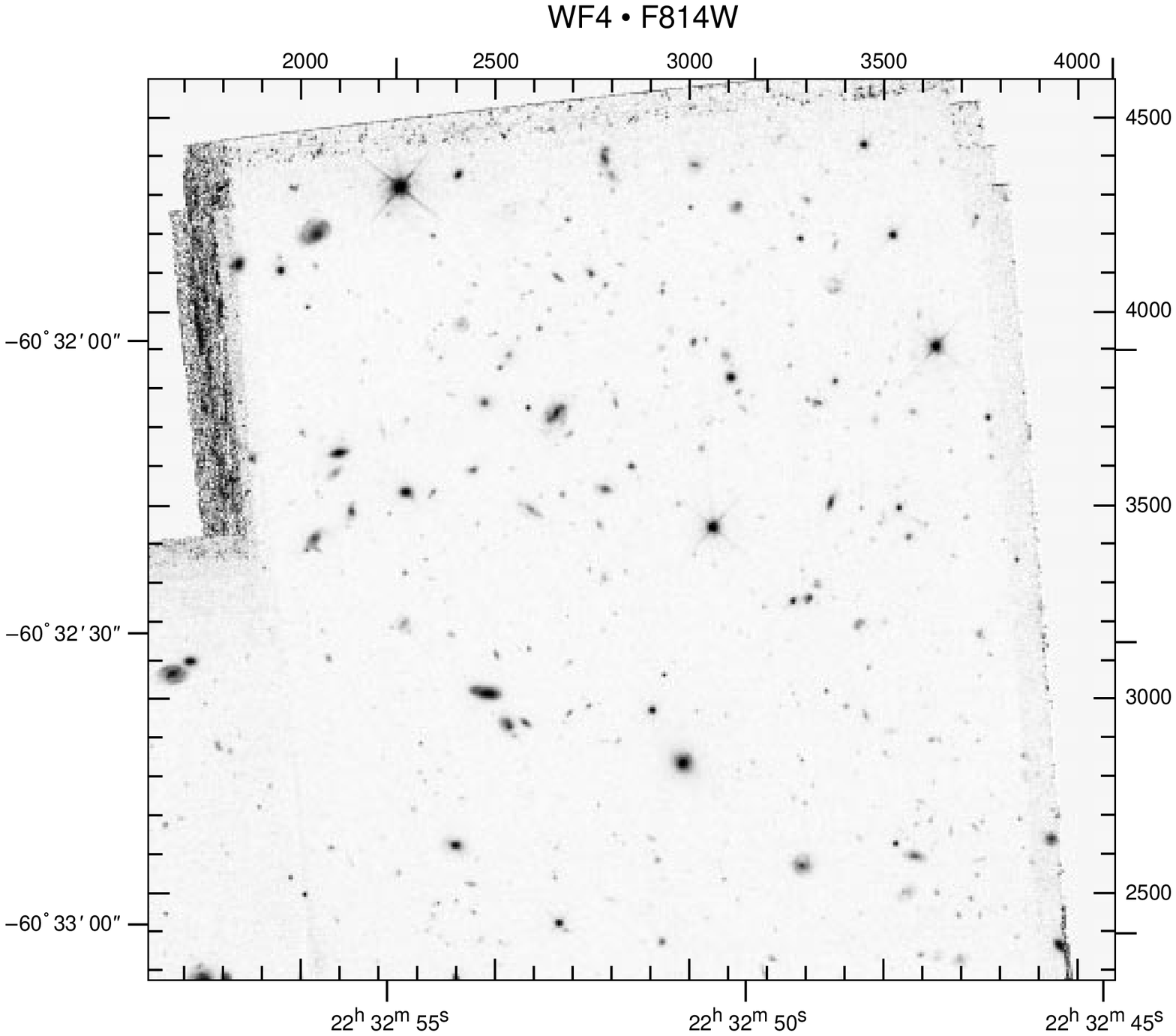,height=6in,bbllx=50pt,bblly=70pt,bburx=560pt,bbury=680pt}}
 \figcaption[Casertano_fig25.ps] { The NW quadrant of the WFPC2 field,
roughly corresponding to the area covered by the Wide Field Camera 4, in
F814W.}
 \end{figure}